\documentclass{article}
\usepackage{graphicx} 
\usepackage{bm,amsmath,color,dsfont, amssymb}
\usepackage[titletoc,toc,title]{appendix}
\usepackage{graphicx}
\usepackage[section]{placeins}
\usepackage{esint}
\usepackage[hidelinks]{hyperref}
\usepackage[justification=centering]{caption}
\usepackage{cleveref}
\usepackage{subcaption}
\usepackage{float}
\usepackage{multicol}
\usepackage{array} 
\usepackage{authblk}
\usepackage{tabularx}
\usepackage{algorithm, algorithmic}
\usepackage{tgtermes} 
\usepackage{relsize}
\usepackage{wrapfig}
\usepackage{placeins}
\usepackage{siunitx}
\usepackage{bm}
\usepackage{xcolor}
\usepackage[
backend=biber,
style=nature,
]{biblatex}
\usepackage{booktabs}
\usepackage{multirow}

\newcommand{\eps}{\varepsilon}

\DeclareMathOperator{\EX}{\mathbb{E}}

\begin{document}

\title{The Subtle Interplay between Square-root Impact, \\ Order Imbalance \& Volatility: {A Unifying Framework}}

\author[1, 2, 3]{Guillaume Maitrier \thanks{Corresponding author: guillaume.maitrier@hotmail.fr, second author : jean-philippe.bouchaud@cfm.com}}
\author[4, 2, 5]{Jean-Philippe Bouchaud}

\affil[1]{\textit{LadHyX UMR CNRS 7646, École polytechnique, 91128 Palaiseau, France}}
\affil[2]{\textit{Chair of Econophysics and Complex Systems, École polytechnique, 91128 Palaiseau, France}}
\affil[3]{
\textit{BNP Paribas Global Markets, 20 Boulevard des Italiens, 75009 Paris, France}}
\affil[4]{\textit{Capital Fund Management, 23-25 Rue de l'Université, 75007 Paris, France}
}
\affil[5]{\textit{Académie des Sciences, Paris 75006, France}}

\maketitle

\begin{abstract}
In this work, we aim to reconcile several apparently contradictory observations in market microstructure: is the famous ``square-root law'' of metaorder impact, which decays with time, compatible with the random-walk nature of prices and the linear impact of order imbalances? Can one entirely explain the volatility of prices as resulting from the flow of uninformed metaorders that mechanically impact them? We introduce a new theoretical framework to describe metaorders with different signs, sizes and durations, which {\it all} impact prices as a square-root of volume but with a subsequent time decay. We show that, as in the original propagator model, price diffusion is ensured by the long memory of cross-correlations {\it between} metaorders. 
In order to account for 
the effect of strongly fluctuating volumes $q$ of individual trades, we further introduce two $q$-dependent exponents, which allow us to describe how the moments of generalized volume imbalance and the correlation between price changes and generalized order flow imbalance scale with $T$. We predict in particular that the corresponding power-laws depend in a non-monotonic fashion on a parameter $a$, which allows one to put the same weight on all child orders or to overweight large ones, a behaviour that is clearly borne out by empirical data. We also predict that the correlation between price changes and volume imbalances should display a maximum as a function of $a$, which again matches observations. Such noteworthy agreement between theory and data suggests that our framework correctly captures the basic mechanism at the heart of price formation, namely the average impact of metaorders. We argue that our results support the ``Order-Driven'' theory of excess volatility, and are at odds with the idea that a ``fundamental'' component accounts for a large share of the volatility of financial markets. 
\end{abstract}

\textit{Keywords: Market microstructure, price impact, square-root law, order flow dynamics, diffusivity puzzle, physics of financial markets.}

\newpage
\tableofcontents
\newpage
\section{Introduction}

{\it Price impact} refers to the fact that buyers push the price up and sellers push the price down \cite{bouchaud2010impact, bouchaud2018trades}. The traditional {\it Efficient Market} interpretation of this empirical fact is that buyers and sellers are on average informed, and lo and behold, the price moves according to their prediction \cite{hasbrouck1991measuring, hasbrouck2007empirical}. This scenario is at the heart of the celebrated Kyle model for price impact \cite{kyle1985continuous}. 

Another, very different interpretation of price impact is that it is a purely statistical reaction of the market to incoming order flow, where information plays little role. Prices move simply because people trade, whatever the reason they are trading, and volatility is the result of people randomly buying and selling. This is the {\it Order-Driven} view of markets, explicitly spelled out in \cite{bouchaud2018trades}, chapter 20 -- but see also \cite{bouchaud2006random, hopman2007supply, bouchaud2010impact} and in a different setting, \cite{gabaix2021search, bouchaud2022inelastic}. In this scenario, there is no ``information revelation'' but rather ``self-fulfilling prophecies'', as emphasized in \cite{bouchaud2009markets, van2024ponzi}.

Of course, reality should lie somewhere between these two extremes. There are surely some informed trades, and news do impact prices, but there is also overwhelming evidence for the presence of ``noise traders'', excess trading and excess volatility in financial markets, see e.g. \cite{black1986noise, odean1999investors, shiller1981stock, leroy2006excess, kurth2025revisiting}. Direct empirical estimates suggest that informed trades are a minority (see \cite{bouchaud2003fluctuations} and \cite{bouchaud2018trades}, chapter 16). While we are convinced that the Order-Driven view is a much closer approximation to the dynamics of markets, there are several empirical loose ends that need to be tied up. A major issue is how to reconcile two prominent stylized facts of market microstructure, namely, (i) the long-range memory in order signs (i.e. $+1$ for buy orders and $-1$ for sell orders) and (ii) the ubiquitous square-root law of market impact (that governs the average price move induced by the execution of a sequence of orders) with the random walk nature of prices, with a volatility $\sigma$ that is directly proportional to the amplitude of the square-root law. More precisely, the square-root law states that the average price impact $\mathcal{I}$ of a {\it metaorder} of total volume $Q$ is given by 
\begin{equation}\label{eq:sqrt}
    \mathcal{I}(Q) = Y \sigma \sqrt{\frac{Q}{\phi}},
\end{equation}
where $Y$ is a $O(1)$ numerical coefficient and $\phi$ is the average flow of orders executed in the market per unit time -- see e.g. \cite{loeb1983trading, grinold2000active, almgren2005direct, toth2011anomalous, bacry2015market, donier2015million, toth2016square, bucci2018slow, AQR, sato2024does, maitrier2025generatingrealisticmetaorderspublic} and \cite{bouchaud2018trades, webster2023handbook} for reviews. 

Eq. \eqref{eq:sqrt} may look familiar, and in fact trivial: superficially, it states that price changes (i.e. $\mathcal{I}(Q)$) grow as the square-root of execution time $T$, i.e. the law of random walks, {\it provided} one assumes that $T$ and $Q$ are proportional. But, as argued in \cite{bucci_impact_vol2019}, such an argument is completely misleading: not only $\mathcal{I}(Q)$ is an average price change and not a standard deviation; but also $\mathcal{I}(Q)$ is found to depend {\it only} on the quantity executed $Q$ and {\it not} on execution time $T$. Furthermore, it is known that the impact of metaorders {\it decays}, on average, after the end of the execution period (e.g. \cite{moro2009market, brokmann2015slow, bucci2018slow}). Note that all these features are at odds with the Kyle model that predicts linear and permanent impact, resulting from information revelation \cite{kyle1985continuous}.\footnote{Related to this discussion, see also \cite{saddier2024bayesian}.}

We are thus confronted with three interrelated but separate problems: 
\begin{itemize}
    \item (a) What is the basic mechanism that explains the square-root law, Eq. \eqref{eq:sqrt}, and its surprising universal character?
    \item (b) Can the volatility of prices $\sigma$ be explained only in terms of the impact of intertwined, possibly uninformed metaorders? Or is it impact that is somehow slaved to some ``fundamental'' volatility? 
    \item (c) Can one reconcile the {\it square-root} law for metaorders with a {\it linear} relation between average price changes and order flow imbalance?
\end{itemize} 

Although many theories have been proposed to explain the square-root law, there is no consensus on the issue. The Latent Liquidity Theory proposed in \cite{donier2015fullyconsistentminimalmodel} (see also \cite{donier2016walras, bouchaud2018trades}) seems to capture many features observed in data but seems inconsistent with others, in particular those reported in the recent paper of Maitrier et al. \cite{maitrier2025double} using the Tokyo Stock Exchange detailed database. We will not attempt to dwell further on this particular issue here but accept it as an incontrovertible empirical fact, still waiting for a fully convincing explanation. We will rather focus on points (b) and (c) above: knowing that the duration of metaorders is power-law distributed, can one reconcile the square-root impact law with the volatility of markets and with a locally linear aggregate impact law? 

\section{Main contributions, key results and roadmap}\label{sec:contrib_roadmap}

This paper proposes a unified, metaorder-based framework that connects four empirical pillars of market microstructure:
\emph{(i)} the square-root law of metaorder impact,
\emph{(ii)} the post-execution decay of impact,
\emph{(iii)} the long memory of order flow,
and \emph{(iv)} diffusive prices.
The objective is not to derive the square-root law from first principles, but to study its consequences when combined with heavy-tailed metaorder durations, heterogeneous trade sizes, and (crucially) correlations \emph{between} metaorders.

\vspace{0.2cm}
\noindent
\textbf{Main results.} The technical developments of the paper lead to three takeaways:

\begin{enumerate}
    \item \textbf{Diffusion with transient square-root impact requires cross-metaorder correlations.}
    When metaorder impact is transient (decaying after execution), the heavy-tailed distribution of metaorder durations
    \emph{alone} leads to mean-reversion, i.e. subdiffusive impact-induced variance (Section~\ref{Part:ImpactDiffusivity}).
    We show that price diffusion is restored when the \emph{signs of different metaorders} are long-range correlated,
    $\mathbb{E}[\varepsilon(t)\varepsilon(t+\tau)]\sim\tau^{-\gamma_\times}$, because the resulting contribution to price variance
    scales as $T^{2-\gamma_\times-2\beta}$ (Eq.~\eqref{eq:vol_long_range_od}).
    This yields diffusion for $\gamma_\times\simeq 1-2\beta$, which is empirically falsifiable using, e.g. the Tokyo Stock exchange data of \cite{sato2023inferring, sato2024does}.

    \item \textbf{Volume heterogeneity induces non-trivial, $a$-dependent imbalance scaling.}
    To bridge the gap between \emph{sign} $\varepsilon$ imbalance and \emph{volume} $q$ imbalance, we introduce the generalized imbalance
    $I_T^a=\sum_{t\in[0,T]}\varepsilon_t (q_t)^a$ (with exponent $a=0$ and $a=1$ as special cases).
    Motivated by data, we allow the metaorder-duration tail exponent to depend on the typical child order size, $\mu_q$, in such a way that
    large-$q$ orders are empirically less persistent.
    This mechanism predicts an $a$-dependent crossover in the scaling of the moments of $I_T^a$
    (Eqs.~\eqref{eq:scaling_Ia_2}--\eqref{eq:scaling_Ia_n}), confirmed empirically (Fig.~\ref{fig:empiricalIa}).

    \item \textbf{Reconciling square-root metaorder impact with locally linear aggregated impact observables.}
    The above ingredients yield testable predictions for covariances and correlations between returns $\Delta_T$ and generalized imbalances $I_T^a$.
    In particular, the effective scaling exponent of $\mathbb{E}[\Delta_T \cdot I_T^a]$ as a function of $T$ is predicted to be
    \emph{non-monotonic} in $a$ (Section~\ref{sec:SQLvsAgg}, Eqs.~\eqref{eq:scaling_IDelta_q}--\eqref{eq:scaling_IDelta_q2}),
    and the correlation coefficient $R_a(T)$ is predicted to be \emph{hump-shaped} as a function of $a$
    (Section~\ref{sec:correlation_coef}, Eq.~\eqref{eq:R_a_vs_a}).
    Both features are observed in the data (Figs.~\ref{fig:cov_vs_T_and_a}--\ref{fig:corr_with_fit}).
\end{enumerate}

\vspace{0.2cm}
\noindent
\textbf{Relation to the ``double square-root law''.}
Recent evidence suggests that (i) peak impact follows a square-root in executed volume and (ii) the instantaneous impact of a child order
scales as $\sqrt{q}$ \cite{maitrier2025double}.
Our framework is designed to be consistent with this empirical input: we take $\theta(q)\propto \sqrt{q}$
(Sections~\ref{Part:ImpactDiffusivity}--\ref{subsec:diffusion_correlation_meta}), and study how such a microscopic scaling propagates to
mesoscopic observables (diffusion, covariances, correlations) once metaorders overlap and relax.

\vspace{0.2cm}
\noindent
\textbf{Implications for volatility generation (order-driven vs.\ fundamental).}
Because diffusion can be obtained from the superposition of (possibly correlated) metaorders with transient impact,
the model naturally predicts an impact amplitude proportional to volatility (Eq.~\eqref{eq:Y_od}).
We also derive how an additional exogenous ``fundamental'' component would affect covariances/correlations
(Section~\ref{subsec:pref_informed}), yielding scalings that are not supported by the observed $a$-dependence.
This provides empirical evidence that a substantial fraction of volatility is mechanically generated by trading activity.

\subsection*{Roadmap}

To address questions (b) and (c) quantitatively, we first introduce in Section~\ref{Part:OrderFlow} a continuous-time description of the order flow as an overlapping population of metaorders with heterogeneous signs, durations and child-order sizes. Throughout, we take as an empirical input that each metaorder generates a \emph{square-root} peak impact in executed volume, followed by a \emph{decay} after execution.

In Section~\ref{Part:OrderFlowImbalance}, we study the statistics of (generalized) order-flow imbalances over windows of size $T$. In particular, to account for the strong heterogeneity of trade sizes $q$, we introduce a family of imbalances $I_T^a$ that interpolates between sign imbalance ($a=0$) and volume imbalance ($a=1$). This leads us to introduce two $q$-dependent exponents (empirically motivated and estimated from the data) that govern both the persistence of order flow and the decay of impact. These ingredients yield non-trivial scaling predictions for the moments of $I_T^a$ as a function of $T$.

In Section~\ref{Part:ImpactDiffusivity}, we then turn to the impact--diffusivity puzzle: how can transient, concave (square-root) impact be compatible with diffusive prices? We show that, in close analogy with the original propagator mechanism, diffusion is restored by the long memory of \emph{cross-correlations between distinct metaorders}. This identifies metaorder cross-correlations as a crucial ingredient of an order-driven volatility mechanism when impact decays.

Finally, in Sections~\ref{sec:SQLvsAgg}--\ref{sec:correlation_coef}, we connect the above building blocks to observables that can be measured directly from trade and price data, namely covariances and correlations between returns and generalized imbalances. We predict that the relevant scaling exponents depend \emph{non-monotonically} on the weighting parameter $a$ (which interpolates between equal weight per trade and overweighting large trades), and that the correlation between returns and imbalances is \emph{hump-shaped} as a function of $a$. Both predictions are borne out empirically.

We conclude by discussing the implications of these results for volatility generation: the observed scalings and $a$-dependences are naturally explained when volatility is predominantly impact-driven, and are hard to reconcile with the hypothesis that an exogenous ``fundamental'' component accounts for a large fraction of price fluctuations. A simulation-based companion paper~\cite{maitrier2025subtle} broadly confirms the theoretical results presented here and implements the framework as an artificial market generator.

\section{A Continuous Time Description of the Order Flow}\label{Part:OrderFlow}

\subsection{Model set-up}

We posit that between $t$ and $t+{\rm d}t$ and with probability $\nu {\rm d}t$ a new metaorder of random sign $\varepsilon(t) = \pm 1$ and duration $s(t)$ is initiated. The volume of child orders is $q$ (which might itself be random, see below), and during execution the probability that one of them gets executed is $\varphi {\rm d}t$, independently of the size $q$. 

The total size of the metaorder is thus $Q = q \varphi s + O(\sqrt{s})$. The probability density of durations $s$ is denoted $\Psi(s)$, which will typically has a power-law tail $\Psi(s) \propto s^{-1-\mu}$, such that the distribution of metaorder sizes $Q$ inherits from this power-law, and decays as $Q^{-1-\mu}$ as suggested by empirical data \cite{farmer2013efficiency, sato2023inferring}. 
We neglect throughout this paper activity fluctuations as well as intraday seasonalities, as these are not crucial for the effects we want to focus on. This means that $\mu, \varphi$ and the average duration $\bar{s}$ are chosen to be time independent.

Such a power-law distribution of metaorder sizes is the basic mechanism proposed by Lillo, Mike and Farmer (LMF) \cite{lillo2005theory} to explain the long memory of order signs, which is known to decay with lag $\tau$ as $\tau^{-\gamma}$ with $0 < \gamma < 1$ \cite{bouchaud2009markets}. Within the LMF model, one has $\gamma = \mu - 1$, a result recently validated in great details by Sato and Kanazawa \cite{sato2023inferring} using data from the Tokyo Stock Exchange. In a later stage, we will allow the exponent $\mu$ to depend on $q$, to account for the fact that large child orders tend to be less autocorrelated that small ones.

We will also allow the sign of different metaorders to be correlated, as indeed observed in data \cite{donier2015million, bucci2020co}. More precisely, we will model the long-term decay of the autocorrelation of signs, $\mathbb{E}[\varepsilon(t)\varepsilon(t+\tau)]$ as a power-law $\tau^{-\gamma_\times}$, with an exponent $\gamma_\times$ {\it a priori} such that $\gamma_\times \geq \gamma$ such not to contradict the LMF hypothesis. 

We start warming up by computing two simple quantities, the total number of active metaorders and the average trading volume within a window of duration $T$. We will then turn to the distribution of volume imbalance in windows of different sizes. 

\subsection{Average number of metaorders}

The total number of metaorders $N_T$ that are active between $t=0$ and $t=T$ is given by 
\begin{equation}
    N_T = \int_{-\infty}^T {\rm d}N_t \,\, \mathbb{I}(t+s(t) > 0),
\end{equation}
where ${\rm d}N_t = 0$ if there is no new metaorder initiated between $t$ and $t+{\rm d}t$
and ${\rm d}N_t = 1$ otherwise. This equation means that to be active in $[0,T]$, it must start before $t=T$ and end at least after $t=0$.

The average over the probability of initiating metaorders and over their duration gives
\begin{equation}
    \overline{N}_T = \nu \int_{-\infty}^T {\rm d}t \int_0^\infty {\rm d}s \, \Psi(s) \, \mathbb{I}(t+s > 0) = \nu (T + \bar{s}),  
\end{equation}
where we assume henceforth that the average size of metaorders is finite, i.e. 
$\bar{s} := \int_0^\infty {\rm d}s \,s \Psi(s) < +\infty$, which is tantamount to $\mu > 1$. Hence, for large $T \gg \bar{s}$, one finds $\overline{N}_T \approx \nu T$, as expected. 

In the following, we will always make averages over the metaorder initiation process, and often replace ${\rm d}N_t$ by $\nu {\rm d}t$ whenever possible. When comparing with empirical data, we will work in trade time $N_T$ but still call this quantity $T$. Translating our results is real time is, however, non-trivial because, as is well known, the activity rate $\nu$ shows strongly intermittent dynamics (often modeled using Hawkes processes, see e.g. \cite{bouchaud2018trades}, chapter 9) on top of a U-shaped intraday pattern.

\subsection{Average trading activity and trading volume}

Second warm-up question: what is the total activity $A_T$ and total trading volume $V_T$ executed between $t=0$ and $t=T$?
In the following we assume that all child orders have the same size and denote $\kappa := q \varphi \nu$, so that activity and trading volume are simply related by $\overline{V}_T=q \overline{A}_T$. More generally, the following results holds with $\kappa = \bar{q} \varphi \nu$.

There are two terms, corresponding to metaorders initiated within the period $[0,T]$ or before $t=0$ that are still active in $[0,T]$. We denote these two terms as $V_T^1$ and $V_T^2$, with
\begin{equation}
    V_T^1 = \int_0^T {\rm d}N_t \left[\mathbb{I}(t+s(t) > T) q \varphi (T-t) + \mathbb{I}(t+s(t) < T) q \varphi s\right],
\end{equation}
which after averaging over ${\rm d}N_t$ gives
\begin{equation}
    V_T^1 = \kappa\int_0^T {\rm d}t \left[\mathbb{I}(t+s(t) > T) (T-t) + \mathbb{I}(t+s(t) < T) s\right].
\end{equation}
Similarly, for $V_T^2$ we get
\begin{equation}
    V_T^2 = \kappa \int_{-\infty}^0 {\rm d}t  \left[\mathbb{I}(t+s(t) > T) T + \mathbb{I}(0 < t+s(t) < T) (s(t)+t)\right]
\end{equation}
Now let us compute the average over duration $s$, given by
\begin{equation}
    \overline{V}_T^1 = \kappa \int_0^T {\rm d}t \int_0^\infty {\rm d}s \, \Psi(s) \left[\mathbb{I}(t+s > T) (T-t) + \mathbb{I}(t+s < T) s\right]
\end{equation}
and 
\begin{equation}
    \overline{V}_T^2 = \kappa \int_{-\infty}^0 {\rm d}t \int_{0}^\infty {\rm d}s \, \Psi(s) \left[\mathbb{I}(t+s > T) T + \mathbb{I}(0 < t+s < T) (s+t)\right]
\end{equation}
Carefully taking the derivative with respect to $T$ one finds that the total average executed volume $\overline{V}_T = \overline{V}_T^1+\overline{V}_T^2$ is given, for large $T$, by 
\begin{equation}
    \overline{V}_T = \kappa \bar{s} T \approx q \varphi \bar{s} \, \overline{N}_T,
\end{equation} 
i.e. the average volume per metaorder $\overline{Q} = q \varphi \bar{s}$ times the average number of metaorders $\overline{N}_T$. So in this model the average volume {\it flow} per unit time is 
$\phi := \nu \overline{Q} = \kappa \bar{s}$. Note that for large enough $T$ it is dominated by $V_T^1$. 

The average {\it activity} (i.e. number of trades per unit time) is given by $\nu \varphi \bar{s}$. We will denote its inverse $\tau_0 := (\nu \varphi \bar{s})^{-1}$, which is the average time between two trades. Finally, note that the average number of child orders per metaorder is $\bar{n} := \varphi \bar{s}$.

\section{Order Flow Imbalance}\label{sec:scaling_imbalances}\label{Part:OrderFlowImbalance}

In this section we compute, within our model, the statistics of the flow imbalance during periods of duration $T$. In the case where all trades have the same size $q$, volume imbalance is trivially proportional to sign imbalance. It turns out that due to the heavy tail in metaorder durations, sign imbalance scales anomalously with $T$ and has non-Gaussian fluctuations, even when the signs of metaorders are independent. 

When single trade volumes are themselves strongly fluctuating, the statistics of volume imbalance can be very different from those of sign imbalance, a feature we actually observe on data and reproduce within our framework -- see section \ref{sec:volume_fluct} below.  

\subsection{Sign Imbalance}

The sign imbalance $I^0_T$ in an interval of size $T$ is given by a sum of two contributions, as for the traded volume above: 
\begin{equation}
    I^0_{T,1} = \varphi \int_0^T \varepsilon(t) {\rm d}N_t \left[\mathbb{I}(t+s(t) > T) (T-t) + \mathbb{I}(t+s(t) < T) s\right],
\end{equation}
and 
\begin{equation}
    I^0_{T,2} = \varphi  \int_{-\infty}^0 \varepsilon(t) {\rm d}N_t  \left[\mathbb{I}(t+s(t) > T) T + \mathbb{I}(0 < t+s(t) < T) (s(t)+t)\right].
\end{equation}
Because $\mathbb{E}[\varepsilon(t)]=0$, these terms are of mean zero. In this subsection and the next, we assume metaorders to be independent, in particular one has $\mathbb{E}[\varepsilon(t) {\rm d}N_t \varepsilon(t') {\rm d}N_{t'}]= \delta(t-t') {\rm d}N_t$.  

The sign imbalance variance is then given by $\varphi^2 \nu$ times
\begin{align}
     &\int_0^T {\rm d}t \int_0^\infty {\rm d}s \, \Psi(s) \left[\mathbb{I}(t+s > T) (T-t) + \mathbb{I}(t+s < T) s\right]^2  \nonumber \\+ &\int_{-\infty}^0  {\rm d}t \int_{0}^\infty {\rm d}s \, \Psi(s) \left[\mathbb{I}(t+s > T) T + \mathbb{I}(0 < t+s < T) (s+t)\right]^2
\end{align}
We note that because the indicator functions are non-overlapping, all cross-products are zero. 
Taking a derivative with respect to $T$ of the previous expression and denoting the result $D_2$ we get
\begin{equation}
   D_2=  \int_0^T {\rm d}s \,s^2 \Psi(s) + 2T \int_T ^\infty {\rm d}s \, s\Psi(s) - T^2 \int_T ^\infty {\rm d}s \, \Psi(s)
\end{equation}
Suppose for definiteness that 
\begin{equation}\label{eq:psi_s}
 \Psi(s) = \frac{\mu s_0^\mu}{s^{1 + \mu}}\, \mathbb{I}(s>s_0), \qquad  1 < \mu < 2,  
\end{equation} 
corresponding to a sign autocorrelation function decaying as $\tau^{-\gamma}$ with $\gamma = \mu -1 < 1$, as found in the data \cite{lillo2005theory,sato2023inferring}. Then the previous expression becomes:
\begin{equation}\label{eq_D0}
    D_2 = \frac{2}{(2-\mu)(\mu-1)} s_0^\mu T^{2-\mu}
\end{equation}
Hence in this case the variance of the sign imbalance is given by
\begin{equation} \label{eq:Sigma}
    \Sigma^2 = \frac{2 \varphi^2 \nu}{(3 - \mu)(2-\mu)(\mu-1)} s_0^\mu T^{3-\mu},
\end{equation}
i.e. a growth faster than $T$ but slower than $T^2$. Note that when $\mu \nearrow 2$, one smoothly recovers the expected result for a short-range correlated order flow, namely $\Sigma^2 \propto T$. 

One can also compute the fourth moment of the sign imbalance. Focusing on the $I^0_{T,1}$ contribution, one finds
\begin{align}
     \mathbb{E}[(I^0_{T,1})^4]=\varphi^4 \nu &\int_0^T {\rm d}t \int_0^\infty {\rm d}s \, \Psi(s) \left[\mathbb{I}(t+s > T) (T-t) + \mathbb{I}(t+s < T) s\right]^4,  
\end{align}
and taking the derivative with respect to $T$ yields
\begin{align}
   D_4 = \int_0^T {\rm d}s \,s^4 \Psi(s) + T^4 \int_T ^\infty {\rm d}s \, \Psi(s).
\end{align}
When $\mu < 4$, this behaves as $T^{4 - \mu}$, so that the kurtosis of the sign imbalance distribution behaves as $T^{5-\mu}/(T^{3 - \mu})^2 \sim T^{\mu - 1}$ which {\it grows} with $T$! An important consequence is that within our model the sign imbalance {\it does not become Gaussian} for large $T$.  

Generalizing to the $2n$-th moment, one finds that it grows with $T$ like $T^{2n+1 -\mu}$ when $\mu < 2n$. This suggests that when $\mu < 2$, the sign imbalance converges at large $T$ towards a {\it truncated Lévy distribution} of index $\mu$ for the rescaled variable $I^0/T^{1/\mu}$, where the truncation takes place for $|I^0| = T$ (see \cite{Wyart_2003} for a very similar calculation, and the Appendix of \cite{moran2024revisitinggranularmodelsfirm} for a proof). Indeed, one can check that the moments of such a truncated Lévy distribution scale with $T$ exactly as above. We will test this prediction in section \ref{sec:volume_emp}.

\subsection{Generalized Volume Imbalance}\label{sec:volume_fluct}

One can generalize the calculation to the volume imbalance, or in fact to any power $a$ of the individual traded volume, $I^a(T)$, given again by the sum two contributions: 
\begin{equation}
    I^a_{T,1} =  \int_0^T {\rm d}N(t) \varepsilon(t) q^a(t)\left[\mathbb{I}(t+s(t) > T) (T-t) + \mathbb{I}(t+s(t) < T) s\right],
\end{equation}
and 
\begin{equation}
    I^a_{T,2} =  \int_{-\infty}^0 {\rm d}N(t) \varepsilon(t) q^a(t)  \left[\mathbb{I}(t+s(t) > T) T + \mathbb{I}(0 < t+s(t) < T) (s(t)+t)\right].
\end{equation}
Note that $a=0$ corresponds to sign imbalances and $a=1$ to volume imbalances. Obviously, if $q(t)=q$ at all times, all these imbalances are equal, up to a trivial factor, to the sign imbalance computed in the previous section. In the following, we will assume that metaorders differ not only by their duration $s$ but also by the size of their child orders, with a joint distribution that we denote as $\Psi_q(s) \Xi(q)$. Inspired by empirical data, we posit that metaorders that execute with larger child order sizes $q$ still have a power-law distributed duration $s$, but with a tail exponent $\mu_q$ that increases with $q$ -- i.e. have a thinner tail. More precisely, the conditional distribution $\Psi_q(s)$ is of the form:
\begin{equation}\label{eq:muq}
    \Psi_q(s) = \frac{\mu_q s_0^{\mu_q}}{s^{1+\mu_q}}, \qquad \mu_q= \mu_1 + \lambda \ell,
\end{equation}
where $\ell=\log q$ and $q=1$ corresponds to the lot size. \footnote{{Intuitively, it makes sense the sign of large child orders should have shorter memory than small ones. Traders who have large quantities $Q$ to execute is likely to trade small lots in order not to reveal information, whereas smaller $Q$ might be possible to execute in a few shots. }}
 
The generalized volume imbalance $I^a(T)$ still has mean zero and variance $\Sigma_a^2$ now given by $I^a_1 + I^a_2$:
\begin{align}
     & \nu \varphi^2 \int_0^T {\rm d}t \int_0^\infty {\rm d}q \, q^{2a} \Xi(q) \int_0^\infty {\rm d}s \, \Psi_q(s) \left[\mathbb{I}(t+s > T) (T-t) + \mathbb{I}(t+s < T) s\right]^2  \nonumber \\+&  \,  \nu \varphi^2 \int_{-\infty}^0  {\rm d}t \int_0^\infty {\rm d}q \, q^{2a} \Xi(q) 
 \int_{0}^\infty {\rm d}s \, \Psi_q(s) \left[\mathbb{I}(t+s > T) T + \mathbb{I}(0 < t+s < T) (s+t)\right]^2
\end{align}
Consider the $I^a_1$ contribution (the $I^a_2$ contribution does not change the conclusion below):
\begin{align}
     \Sigma_{a,1}^2 = \int_0^T {\rm d}u \int_0^\infty {\rm d}q \, q^{2a} \Xi(q) \int_0^\infty {\rm d}s \, \Psi_q(s) \left[\mathbb{I}(s > u) u^2 + \mathbb{I}(s < u) s^2  \right] 
\end{align}
The derivative of this quantity with respect to $T$ gives
\begin{align}
 \partial_T \Sigma_{a,1}^2 =\int_0^\infty {\rm d}q \, q^{2a} \Xi(q) \left[{T^2}\int_T^\infty {\rm d}s \, \Psi_q(s)  + \int_0^T {\rm d}s \, \Psi_q(s) s^2 \right]
\end{align}
Now assume $T$ is large and define $q_2$ such that $\mu_{q_2}=2$. Metaorders with smaller volumes $q < q_2$ thus have an infinite duration variance ($\mu_q \leq 2$), while larger volumes have a finite variance  ($\mu_q > 2$). The two contributions then read
\begin{equation}
\partial_T \Sigma_{a,1,1}^2 =  \int_0^\infty {\rm d}q \, q^{2a} \Xi(q) T^{2- \mu_q},
\end{equation}
and \footnote{Note that the apparent divergence for $\mu_q=2$ is spurious. In fact, the correct expression should read: $$\int_0^\infty {\rm d}q \, q^{2a} \Xi(q) \int_0^T {\rm d}s \, \Psi_q(s) s^2 =  \int_0^{\infty} {\rm d}q \, q^{2a} \Xi(q) 
 \frac{\mu_q}{2- \mu_q}  \left(T^{2- \mu_q} - s_0^{2- \mu_q}\right),$$ which is finite for all $\mu_q$.}
\begin{equation} \label{sigma_a12}
\partial_T \Sigma_{a,1,2}^2 =  \int_0^\infty {\rm d}q \, q^{2a} \Xi(q) \int_0^T {\rm d}s \, \Psi_q(s) s^2 =  \int_0^{q_2} {\rm d}q \, q^{2a} \Xi(q) 
 \frac{\mu_q T^{2- \mu_q}}{2- \mu_q} + \int_{q_2}^\infty {\rm d}q \, q^{2a} \Xi(q)  \frac{\mu_q}{\mu_q - 2}.
\end{equation}
A convenient mathematical description of the right tail of child order sizes is a log-normal:
\begin{equation}\label{eq:lognorm}
    \Xi(q) = \frac{1}{q \sqrt{2 \pi \sigma_\ell^2}} e^{-\frac{(\ell - m)^2}{2 \sigma_\ell^2}},
\end{equation}
with $\ell := \log q$ and $m$ is the most likely value of $\log q$. One then gets:
\begin{equation}
 \partial_T \Sigma_{a,1,1}^2 \propto  e^{2ma} T^{2- \mu_m} \int_0^\infty {\rm d}\ell \, \, e^{(\ell-m)(2a- \lambda \log T) - (\ell - m)^2/2 \sigma_\ell^2} \propto  e^{2ma + 2 \sigma_\ell^2 a^2} T^{2 - \widetilde \mu(a)}
\end{equation}
with $\mu_m = \mu_1 + \lambda m$ and an effective exponent $\tilde \mu$ that reads
\begin{equation}
    \widetilde \mu(a) = \mu_m + \lambda \sigma_\ell^2 (2a  - \frac12 \lambda \log T)  
\end{equation}
Let us fix a range of $T$ where the data is fitted, and assume for simplicity that we are in a case where $\frac12 \lambda \log T \ll 1$ while $\lambda \sigma_\ell^2 = O(1)$. Then the expression for the effective exponent $\widetilde \mu$  becomes very simple: 
\begin{equation} \label{eq:tildemu_a}
    \widetilde \mu(a) = \mu_m + 2a \lambda \sigma_\ell^2
\end{equation}
So the effective exponent $\widetilde \mu$ increases with $a$, i.e. an exponent $2 - \widetilde \mu$ that decreases with $a$. When $a=0$, the sign correlation is dominated by the most probable volumes $q=e^{m}$ and we recover the previous result with $\widetilde \mu = \mu_m$.

For the contribution $\partial_T \Sigma_{a,1,2}^2$, one has to separate the cases $\ell < \log q_2$ and $\ell > \log q_2$. Since the integral over $\ell$ is dominated by the region $\ell  \approx \ell^* = m + 2a \sigma_\ell^2$, one can use the same expression as above for the first term in Eq. \eqref{sigma_a12}, when $\ell^* < \log q_2$. When $\ell^* > \log q_2$, $\partial_T \Sigma_{a,1,2}^2$ is dominated by the second term and becomes independent of $T$.

Putting everything together, the predictions of this simple model are thus that
\begin{equation} \label{eq:scaling_Ia_2}
    \Sigma_a^2 \propto  e^{2ma + 2 \sigma_\ell^2 a^2} \times \begin{cases}
        & T^{3 - \widetilde \mu(a)}, \qquad \qquad \, \, a < a_c(1):=\frac{2 - \mu_m}{2 \lambda \sigma_\ell^2}; \\
        & T, \qquad \qquad \quad \quad a \geq a_c(1).
    \end{cases}
\end{equation}
In other words, one finds that the variance of the generalized volume imbalance scales anomalously with $T$ when $a$ is small enough (like the sign imbalance considered above), but becomes simply diffusive when $a$ is large. Intuitively, it is because large child orders are much less auto-correlated than small child orders when $\lambda > 0$.\footnote{\label{foot:scaling_Ia} Another mechanism that leads to dependence of the effective exponent of $\Sigma_a^2(T)$ is the presence of power-law tails in the distribution of $q$ that can be a confounding factor. If the tail exponent is equal to $\varpi$ (see section \ref{subsec:volume_dist}), one expects a crossover value $a_c(n)$ given by $\varpi/2n$, i.e. when $\mathbb{E}[q^{2na}]$ diverges. Although such a mechanism may certainly play a role, it comes in parallel with the dependence of $\mu_q$ on $q$.} We will compare these predictions with empirical data in the next section. Although the model is over-simplified, we will see that it captures the data semi-quantitatively. Typically, $a_c(1)$ is found to be around $2$. With $\mu_m = 3/2$, we find $\lambda \sigma_\ell^2 \approx 1/8$, an estimate that will match other observations.

It is interesting to generalize these results to higher moments of the volume imbalance. Extending the calculation above, one finds 
\begin{align}
     \Sigma_{a,1}^{(2n)} = \int_0^T {\rm d}u \int_0^\infty {\rm d}q \, q^{2na} \Xi(q) \int_0^\infty {\rm d}s \, \Psi_q(s) \left[\mathbb{I}(s > u) u^{2n} + \mathbb{I}(s < u) s^{2n}  \right], 
\end{align}
from which one derives the following result 
\begin{equation}\label{eq:scaling_Ia_n}
    \Sigma_{a,1}^{(2n)} \propto 
    \begin{cases} 
        T^{2n + 1 - \mu_m - 2na \lambda \sigma_\ell^2}, & \quad a <  a_c(n); \\
        T, & \quad a \geq a_c(n),
    \end{cases}
\end{equation}
with $a_c(n)=(1 - \mu_m/2n)/\lambda \sigma_\ell^2$.

\subsection{The role of long-range correlations {\it between} metaorders}\label{sec:gamma_m}

It is known that the signs of metaorders initiated by different traders are also correlated, see \cite{donier2015million, bucci2020co}. This may either be due to herding, or more plausibly to different traders following the same signal. As mentioned above, we assume that the sign cross-correlation $\mathbb{E}[\varepsilon(t) \varepsilon(t+\tau)]$ decays as $\Gamma (\tau_0/\tau)^{\gamma_\times}$, whereas the sizes $q$ and $q'$ are remain independent for simplicity.\footnote{One can extend the following calculations to the case where conditional size distribution of a metaorder starting at $t+\tau$, knowing that one metaorder started at $t$ is
\begin{equation} \label{conditional_psi}
    \Psi_q(s'|s,\tau) = \frac{\tau^b}{1 + \tau^b} \Psi_q(s')+ \frac{1}{1 + \tau^b} \frac{1}{s} F(s'/s),
\end{equation}
where $F(.)$ is a certain function and $b$ a new exponent. The following results are unaffected provided $b > 1$.} When $\Gamma=0$, there is no correlation between successive metaorders.

When all order sizes $q$ are equal, the variance of the sign imbalance again contains two terms, one of them reading 
\begin{align}
     &\Gamma \tau_0^{\gamma_\times} \iint_0^T  \frac{{\rm d}u \, {\rm d}u'}{|u - u'|^{\gamma_\times}}\int_0^\infty {\rm d}s \, \Psi(s) \int_0^\infty {\rm d}s' \, \Psi(s') \left[\mathbb{I}(s > u) u + \mathbb{I}(s < u) s\right]\left[\mathbb{I}(s' > u') u' + \mathbb{I}(s' < u') s'\right].
\end{align}
Taking the derivative with respect to $T$ leads to
\begin{align}
     &\Gamma  \tau_0^{\gamma_\times} \int_0^T \frac{{\rm d}u'}{(T - u')^{\gamma_\times}} \int_0^\infty {\rm d}s \, \Psi(s) \int_0^\infty {\rm d}s' \, \Psi(s') \left[\mathbb{I}(s > T) T + \mathbb{I}(s < T) s\right]\left[\mathbb{I}(s' > u') u' + \mathbb{I}(s' < u') s'\right].
\end{align}
The scaling of this expression with $T$ is found to be $T^{1 - \gamma_\times}$ provided $\mu > 1$, i.e. as soon as the mean size of metaorders is finite. Hence, we get an off-diagonal contribution to $\Sigma^2$ that scales as $T^{2 - \gamma_\times}$, which must be compared to the ``diagonal'' contribution (i.e. for $u=u'$ and $s=s'$) given in Eq. \eqref{eq:Sigma}, which scales as $T^{2 - \gamma}$.

In other words, the LMF model \cite{lillo2005theory} that ascribes the main contribution to sign autocorrelation to long metaorders is only valid if such metaorders are not too strongly correlated between themselves, i.e. when
\begin{equation}\label{eq:gamma_m} 
        \gamma_\times \geq \gamma.
\end{equation}
In view of the empirical data supporting the LMF model, we stick to this assumption henceforth. In fact, one can measure $\gamma_\times$ directly (G. Maitrier, unpublished, see also \cite{donier2015million}) suggesting $\gamma_\times \approx \gamma$.

Let us now include volume fluctuations {\it on top of} long-range correlations between the sign of metaorders. Assuming that the sizes $q,q'$ of the child orders of two different metaorders are independent, one finds that the off-diagonal (o.d.) contribution to $\Sigma_a^2$ reads:
\begin{equation} \label{eq:sigma_a_od}
    (\Sigma_a^2)_{o.d.} \propto \Gamma  e^{2ma +  \sigma_\ell^2 a^2} \, T^{2 - \gamma_\times},
\end{equation}
to be compared with Eq. \eqref{eq:scaling_Ia_2}. 

With $\mu_m = 3/2$ and $\gamma_\times = 1/2$, one therefore concludes that as soon as $a > 0$, the $T \to \infty$ behaviour of $\Sigma_a^2$ should, in principle, be dominated by the off-diagonal contribution. However, for $a$ small the two exponents $3-\widetilde \mu$ and $2 - \gamma_\times$ are indistinguishable, and  the cross-over time $T_\times$ beyond which $(\Sigma_a^2)_{o.d.}$ is dominant soon becomes unreachable when $a$ grows. When $a > a_c(1)$ one finds, with $\Gamma = O(1)$
\begin{equation}
    T_\times \approx  e^{2\sigma_\ell^2 a^2}.
\end{equation}
For $a=2$ and $\sigma^2_\ell=1$, this yields $T_\times \sim 10^{4}$ trades, beyond the range of times scales studied below.

\subsection{Empirical observations}\label{sec:volume_emp}

For this analysis (as well as the remainder of the paper), we have chosen four assets for which we have trade-by-trade prices and signed volumes. We have chosen two stocks from the LSE, one small tick stock (LLOY) with a small tick size, such that the average spread-to-tick ratio equal to $\approx 3$. The second is a medium tick stock (TSCO), with an average spread-to-tick ratio equal to $\approx 1.5$. 
We also selected two liquid futures contracts: the SPMINI, with a spread-to-tick ratio of approximately $1.1$, and the EUROSTOXX, a large-tick asset with a ratio close to one ($\approx 1.02$).
For equities, the dataset goes from 2012 to 2015, while for futures, it covers 2016–2018 for the EUROSTOXX and 2022 for the SPMINI.
This selection allows us to cover two major asset classes actively traded in modern markets, a wide range of spread-to-tick ratios, and nearly a decade of market evolution.

\subsubsection{Child volume distribution} \label{subsec:volume_dist}

\begin{figure}[H]
    \centering
    \includegraphics[width =0.7\linewidth]{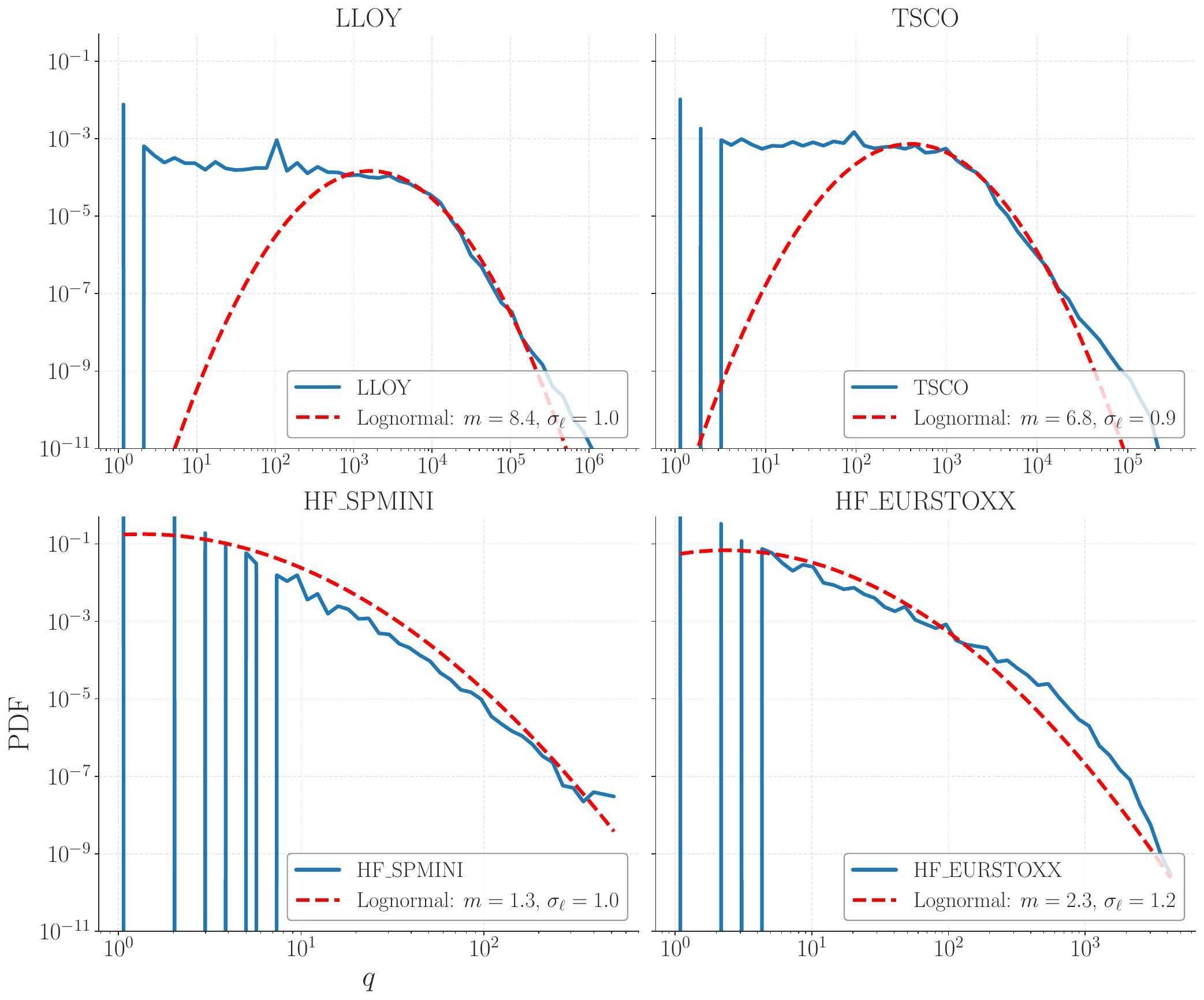}
     \caption{Comparison between the empirical probability distribution functions of the executed volume $q$ for the four selected assets, along with a fitted lognormal distribution in the tail region.}
    \label{fig:lognormal_child_all}
\end{figure}

As discussed in Section~\ref{sec:volume_fluct}, we consider a log-normal distribution for the child order sizes, as defined by Eq.~\eqref{eq:lognorm}. It turns out to be a reasonable approximation of reality for large volumes, see Fig. \ref{fig:lognormal_child_all}, with values of $\sigma_\ell$ reported in the legend, around $1$ for stocks and SPMINI and $1.2$ for EUROSTOXX. A better fit of the tail of the distribution is, arguably, power-law $\propto q^{-1-\varpi}$, with $\varpi$ found to be around $2.1$ -- $2.4$. Such a choice makes the mathematical analysis more cumbersome, and we prefer the log-normal specification for our semi-quantitative discussion of the role of volume fluctuations. Nevertheless, a power-law tail can play a role similar to the coefficient $\lambda$ in $\mu_q$ when it comes to the scaling of $\Sigma^{(2n)}_a$, crossing over to a linear behaviour when $2na \gtrsim \varpi$, see footnote \ref{foot:scaling_Ia}.

\subsubsection{Distribution of sign imbalances}

\begin{figure}[H]
    \centering
    \includegraphics[width=0.6\linewidth]{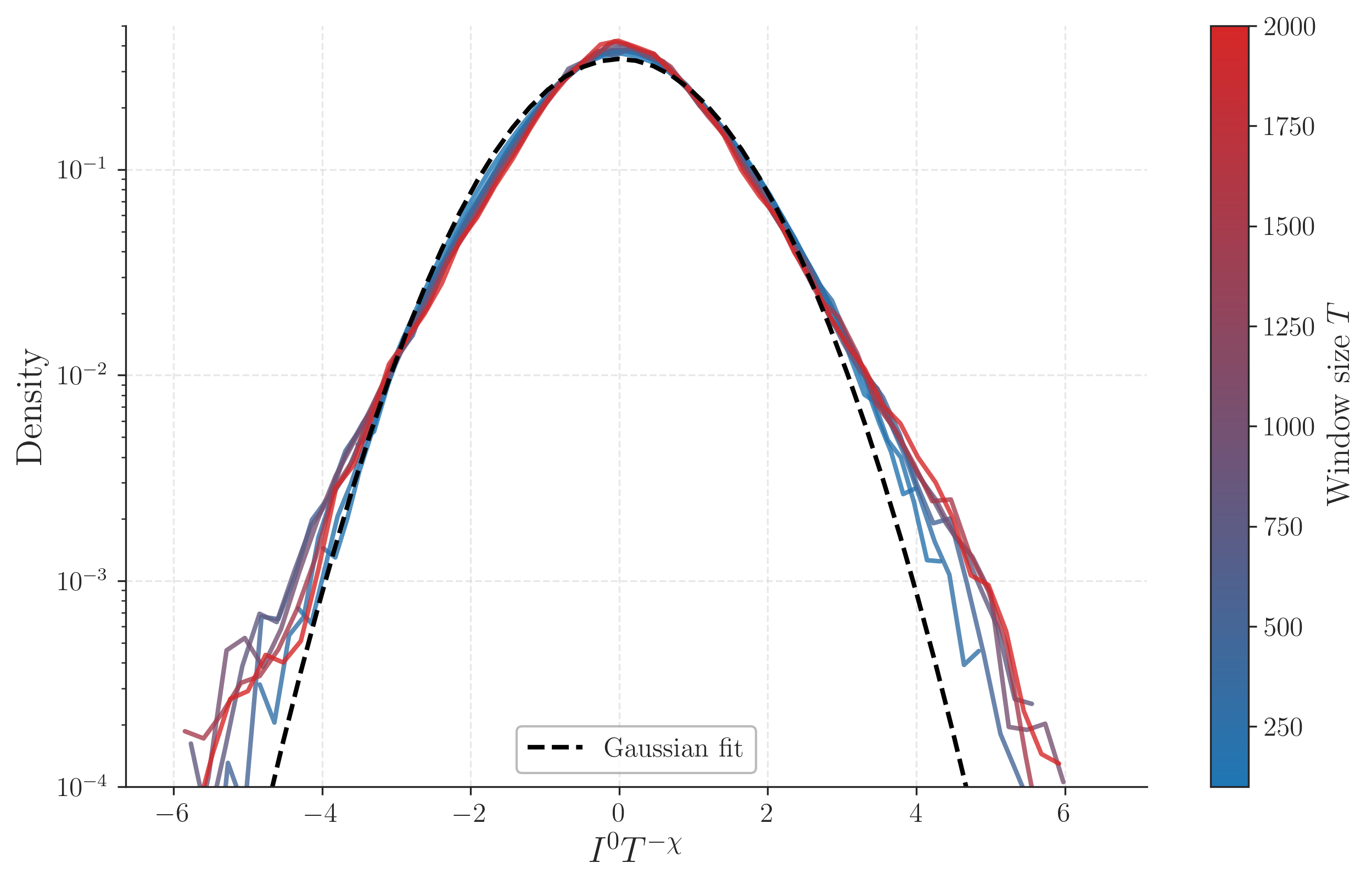}
    \caption{Distribution of sign imbalances $I^0$ as a function of the window size $T$ (measured in number of trades) on EUROSTOXX, between 2016 and 2019. After a proper rescaling by $T^{-\chi}$, with $\chi = 0.72$, distributions nicely collapse onto a single master curve, consistent with findings in \cite{patzelt2018universal}. The value of $\chi$ is not far from the theoretical prediction $\chi = 1/\mu$ with $\mu = 3/2$. The dotted line shows a Gaussian distribution with the same variance as red curve. As expected, the volume imbalance exhibits fat tails.} 
    \label{fig:scaling_imb_EUROSTOXX}
\end{figure}

We show in Fig. \ref{fig:scaling_imb_EUROSTOXX} the distribution of rescaled sign imbalances $I^0 T^{-\chi}$, for different $T$ in trade time and $\chi=0.72$. The theoretical analysis performed in the previous sections predicts that such distributions should collapse when $\chi$ is chosen to be $1/\mu$, where $\mu = 1 +\gamma$ is related to the autocorrelation of the sign of the trades, which gives $\mu \approx 1.4$. Although not perfect, the agreement is quite reasonable, in view of the fact that $\mu$ may actually depend on the size of the child order $q$. The master curve is clearly non-Gaussian, with tails that become fatter as $T$ increases, as expected from our theoretical prediction. 

\subsubsection{Scaling of the generalized volume imbalances}

Our model predicts that the even moments $\Sigma_a^{(2n)}$ of the generalized volume imbalances $I^a$ scale with $T$ with exponents that depend on $a$, see Eq. \eqref{eq:scaling_Ia_n}. In order to test this prediction empirically, we first remark that trade-by-trade data typically exhibit numerous outliers (such as block trades, fat fingers etc...). These outliers can substantially influence the empirical estimation of the diffusion coefficient, particularly for large values of $a$. Consequently, trades quantities were clipped beyond 1\% of the daily volume. 

\begin{figure}[H]
    \centering
    \includegraphics[width=0.7\linewidth]{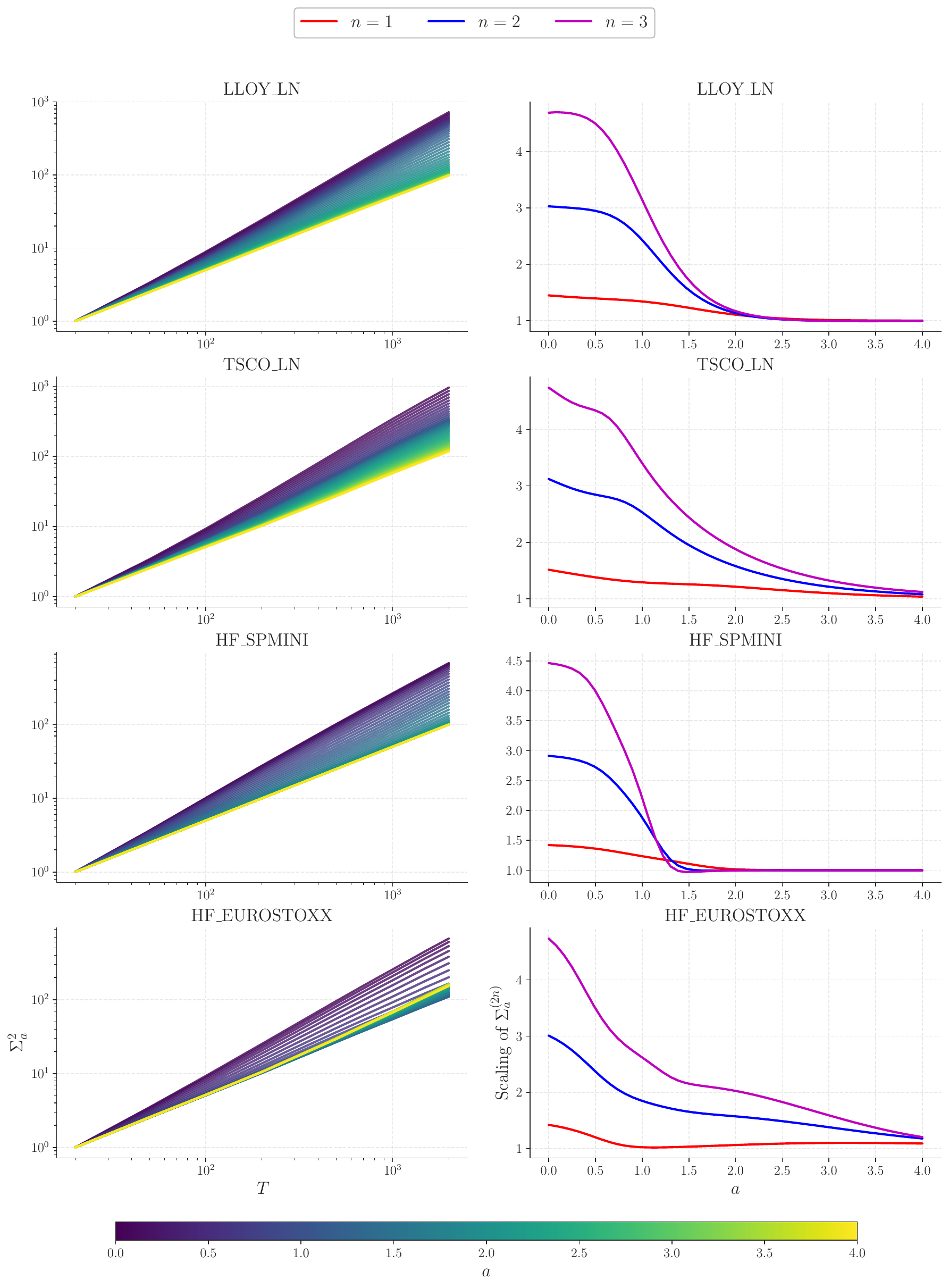}
    \caption{Data for LLOY, TSCO, SPMINI and EUROSTOXX. \textbf{Right column:} scaling of the different moments $\Sigma_{a}^{(2n)}$ as a function of trade time $T$, from which we extract the exponent from a regression of the data in log-log. \textbf{Left column:} scaling exponent as a function of $a$. As predicted by our model, increasing the value of $a$ -- i.e. giving more weight to orders with large volume -- reduces the value of the exponent, which reaches unity for $a > a_c(n)$ with $a_c(1) \approx 2$ for LLOY, TSCO and SPMINI, and $a_c(1) \approx 1$ for EUROSTOXX.}
    \label{fig:empiricalIa}
\end{figure}

The moments $\Sigma_a^{(2n)}$ for $n=1,2,3$ are shown in Fig. \ref{fig:empiricalIa} as a function of $a$. Remarkably, the theoretical predictions qualitatively reproduce the empirical data, in spite of the rather uncontrolled approximations made in the calculations. In particular, we do find that for large enough $a$, all these moments scale proportionally to $T$, whereas super-linear behaviour in $T$ is observed for small $a$, as a consequence of the long memory of order signs. Such behaviour is washed away when we look at large volumes only, i.e. when $a$ is large enough. 

Looking in particular at the curves for $n=1$, we see that $\widetilde \mu(a)$ decreases from $\widetilde \mu(a=0) \approx 3/2$ to $\widetilde \mu(a=a_c) \approx 2$ with $a_c \approx 1$ for large tick EUROSTOXX and  $a_c \approx 2$ for smaller tick LLOY, TSCO and SPMINI. From Eq. \eqref{eq:tildemu_a}, we deduce that $\lambda \sigma_\ell^2 \approx 1/4$ for EUROSTOXX, and 
$\lambda \sigma_\ell^2 \approx 1/8$ for LLOY, TSCO and SPMINI. 

\section{The Impact-Diffusivity puzzle and a generalized propagator model}\label{Part:ImpactDiffusivity}

We now discuss the problem of whether the volatility of price changes can be explained chiefly in terms of the impact of intertwined metaorders of different sizes and signs that get executed in the market, i.e. {\it without} any ``fundamental'' component that makes prices move without any trades. Such fundamental component would merely add a constant to the mechanical (impact induced) component computed below.

We will first recall how price diffusivity and long-memory of order flow are reconciled within the standard ``propagator'' framework, and then show how the issue becomes much more perplexing when the impact of metaorders obeys the square-root law. We postulate that the autocorrelation between different metaorders is responsible for the emergence of price diffusion. Two other possible resolutions are considered in Appendix A, together with their strengths and weaknesses. 

\subsection{Price diffusivity within the standard propagator model}\label{subsec:diffusivity_classic_pro}

We first assume the impact of single orders is given by a \textit{deterministic} propagator model, i.e. the average price change due to an order of volume $q$ executed at time $t=0$ is $\theta(q)$ which then decays as \cite{bouchaud2003fluctuations, bouchaud2018trades}
\begin{equation}
    G(t)=\theta(q) \, \left(\frac{\tau_0}{t + \tau_0}\right)^{\beta},
\end{equation} 
where $\tau_0$ is the average time between two child orders.
The average impact of a single metaorder of volume $Q$, duration $s \gg \tau_0$ and chosen participation rate $\widetilde \varphi$ (such that $Q = q \widetilde \varphi s$) is then given by\footnote{Here we distinguish the participation rate of a specific metaorder, $\widetilde \varphi$, from the average participation rate of the whole market, $\varphi$.}
\begin{equation}\label{eq:prop_impact}
    \mathcal{I}(t \leq s) = \theta(q) \int_0^t {\rm d}t' \, \widetilde \varphi \left(\frac{\tau_0}{t-t'}\right)^{\beta}  = \mathcal{I}_0(q,\widetilde \varphi) t^{1-\beta}, \qquad \mathcal{I}_0(q,\widetilde \varphi):=\frac{\widetilde \varphi  \theta(q) \tau_0^\beta}{1 - \beta}.
\end{equation}
The {\it peak impact} $\mathcal{I}(s)$ reads 
\begin{equation}
    \mathcal{I}(s) = \frac{\widetilde \varphi  \theta(q) s^{1 - \beta} \tau_0^\beta}{1 - \beta} = \frac{\theta(q) q^{\beta - 1}(\widetilde \varphi  \tau_0)^\beta}{1 - \beta} Q^{1 - \beta},
\end{equation}
which reveals one problematic flaw of the propagator model: for $\beta=1/2$, peak impact does not only depend on $Q$, as empirically observed, but also on $q$ and $\widetilde \varphi$. Although one can always choose $\theta(q) \propto q^{1 - \beta}$ to get rid of the $q$ dependence, one is still left with a square-root dependence on the participation rate $\widetilde \varphi$.

\begin{figure}[H]
    \centering
    \includegraphics[width=0.5\linewidth]{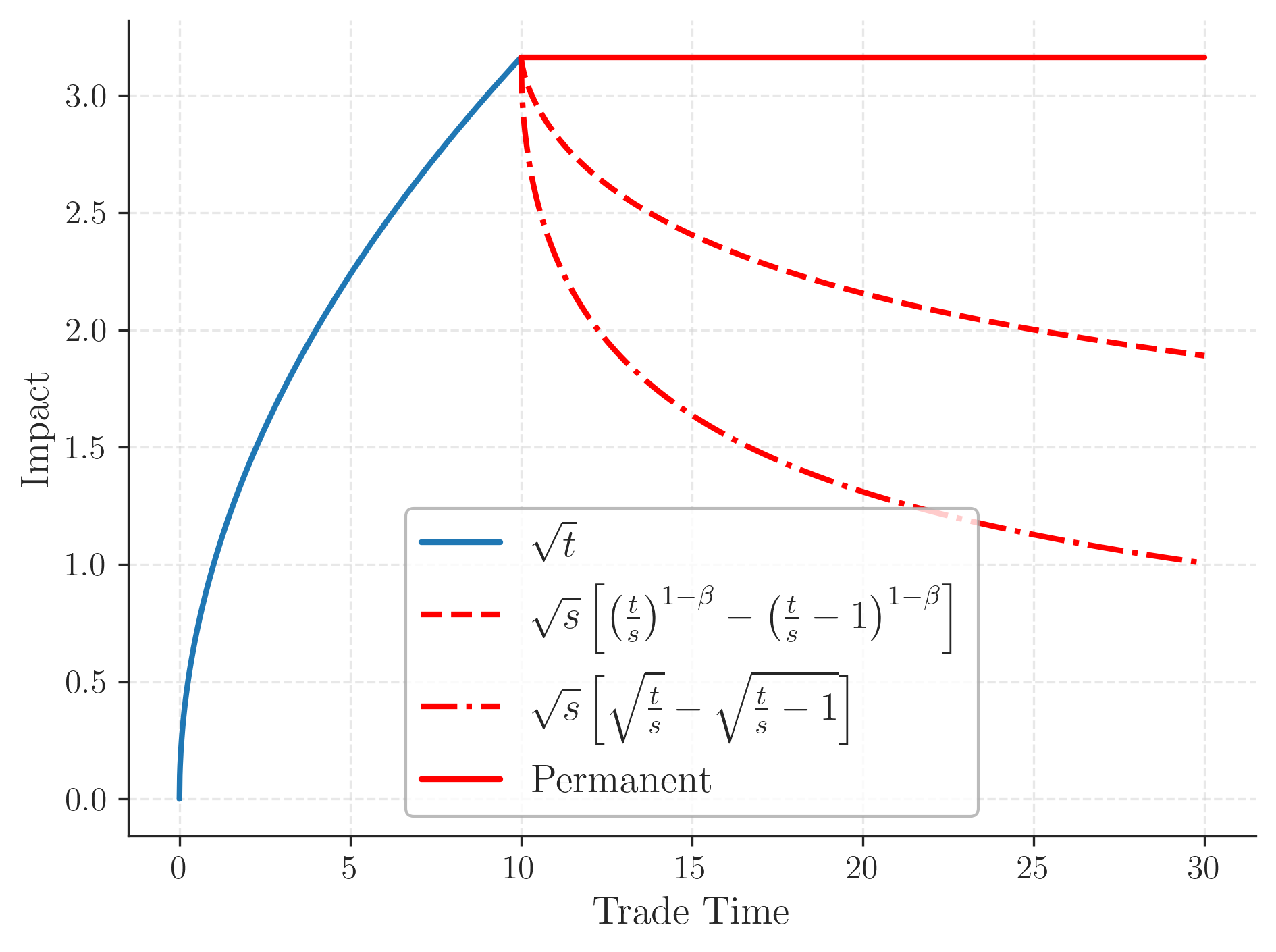}
    \caption{Comparison of the three main theories of impact decay for a metaorder of size $s = 10 $. During the execution phase (blue), impact follows a square-root impact growth, i.e. $\propto \sqrt{t}$, where $t$ denotes the index of the child order \cite{moro2009market, maitrier2025double}. In the post-execution phase (red), impact may either remain permanent (solid line) or decay (dashed). Two distinct decay mechanisms are illustrated: one based on the generalized two-time propagator as $t^{-\beta}$ with $\beta = 0.2$, and the other on the LLOB theory, corresponding to $\beta=\frac12$.}
    \label{fig:different_decays}
\end{figure}

After the end of the metaorder, impact decays (see Fig. \ref{fig:different_decays}) and is given by
\begin{equation}\label{eq:prop_decay}
    \mathcal{I}(t > s) = \theta(q) \int_0^s {\rm d}t' \, \widetilde \varphi\left(\frac{\tau_0}{t-t'}\right)^{\beta}  = \mathcal{I}_0(q,\widetilde \varphi)  \left(t^{1-\beta} - 
    (t-s)^{1- \beta}\right).
\end{equation}
Note that this last expression behaves as $\mathcal{I}(s)(s/t)^\beta$ for $t \gg s$.

For simplicity, we assume for now that all market orders have the same volume $q$, and model the price variation $\Delta_T$ over time $T$ as the superposition of the {\it average} impact of metaorders, neglecting fluctuations that will be considered in section \ref{subsec:impactfluctuation} below. 

Neglecting any ``fundamental'' contributions, price variations are then given by the sum of two terms, $\Delta_{T,1}$ describing the impact of metaorders initiated within $[0,T]$, and $\Delta_{T,2}$ the decaying impact of metaorders initiated before $t=0$:
\begin{equation}
    \Delta_{T,1} = \mathcal{I}_0(q, \varphi)\int_0^T {\rm d}N_t \,\eps(t) \left[\mathbb{I}(t+s > T) (T-t)^{1-\beta} + \mathbb{I}(t+s < T)\left((T-t)^{1-\beta} - (T-t-s)^{1-\beta}\right) \right] 
\end{equation}
and 
\begin{align}
    \Delta_{T,2} &= \mathcal{I}_0(q,\varphi) \int_{-\infty}^0 {\rm d}N_t \, \eps(t)\left[\mathbb{I}(t+s > T) \left((T-t)^{1-\beta}-(-t)^{1-\beta}\right) \right. \nonumber \\
    &\hspace{4cm} \left. + \mathbb{I}(t+s < T)\left((T-t)^{1-\beta} - (T-t-s)^{1-\beta}-(-t)^{1-\beta}\right) \right]
\end{align}
The average of $\Delta_T$ over $\varepsilon$ is of course nil, and its variance is given by two contributions : 
\begin{align}
    \Sigma^2_{T,1} = \mathcal{I}_0^2(q,\varphi)\nu \int_0^T {\rm d}u \,\int_0^\infty {\rm d}s \, \Psi(s)\left[\mathbb{I}(s > u) u^{2(1-\beta)}+ \mathbb{I}(s < u)\left(u^{1-\beta} - (u-s)^{1-\beta}\right)^2 \right] 
\end{align}
and 
\begin{align}
    \Sigma^2_{T,2} &=  \mathcal{I}_0^2(q,\varphi) \nu \int_{T}^\infty {\rm d}u \,\int_0^\infty {\rm d}s \, \Psi(s)\left[\mathbb{I}(s > u) \left(u^{1-\beta}-(u-T)^{1-\beta}\right)^2 \right. \nonumber \\
    &\hspace{4cm} \left. + \mathbb{I}(s < u)\left(u^{1-\beta} - (u-s)^{1-\beta}-(u-T)^{1-\beta}\right)^2 \right] 
\end{align}
All these contributions can be exactly computed for large $T$ when $\Psi(s)$ decays as a power-law $s^{-(1+ \mu)}$, but provided $\mu < 2$ the scaling can simply be obtained by the change of variables $s = xT$, $u = yT$, that yields
\begin{equation}
    \Sigma_T^2 := \mathbb{E}[\Delta_T^2] \propto \mathcal{I}_0^2(q,\varphi) \, T^{3 - 2\beta -\mu}.
\end{equation}
Hence, we see that metaorders contribute to volatility provided $3 - 2 \beta - \mu = 1$. This equality coincides, as expected, with the critical condition $2 \beta = 1 - \gamma$ derived within the propagator model (recall that $\gamma = \mu - 1$). When $\beta < (1-\gamma)/2$, the price is super-diffusive (i.e. $\gg T$), whereas when $\beta > (1-\gamma)/2$, the contribution of the average impact of metaorders to price variance is negligible, i.e. $o(T)$. 

In order to recover the square-root impact law, one should naively set $\beta=1/2$, such that $3 - 2 \beta - \mu = 1 - \gamma$. But the contribution of metaorders to volatility $\Sigma_T^2$ would be then subdominant at long times, unless $\gamma \to 0$ (i.e. an hyper-slow decay of the sign autocorrelation function). Note that the choice $\beta=1/2^-, \gamma = 0^+$ corresponds to the model advocated by Jusselin \& Rosenbaum \cite{jusselin2018noarbitrageimpliespowerlawmarket}, but is difficult to reconcile with the empirically determined value $\gamma \approx 0.5$ \cite{sato2023inferring}. This value of $\gamma$, in turn, means that a square-root impact appears to be unable to generate price diffusion, since in this case $3 - 2 \beta - \mu \approx 0.5 < 1$.

One could then argue that volatility does not primarily come from the average impact of metaorders, but rather from its fluctuations, a possibility that we explore in section \ref{sec:volume_sqrt} below. But in any case, the propagator model with $\beta=1/2$ fails to account for two important stylized facts:
\begin{itemize}
    \item Injecting $\beta=1/2$ into Eq. \eqref{eq:prop_impact}, one finds, as already mentioned above, $\mathcal{I}(s) \propto  \sqrt{\widetilde \varphi \tau_0} \,  \sqrt{Q}$ when $\theta(q) = \sqrt{q}$ (as indeed suggested by the data of \cite{maitrier2025double}). Hence, one recovers the square-root law $\mathcal{I}(Q) = Y \sqrt{Q}$ but with an extra square-root dependence of the prefactor $Y$ on the participation rate $\widetilde \varphi$, when empirical data show that $Y$ is all but independent of $\widetilde \varphi$.
    \item The decay of impact after the end of a metaorder, when fitted with Eq. \eqref{eq:prop_decay}, suggests a value $\beta \approx 0.2 < 1/2$ \cite{bacry2015market, bucci2018slow, maitrier2025double}, i.e. a much faster short time decay and a much slower long time decay than predicted by $\beta=1/2$. 
\end{itemize}

We conclude that the propagator model, even with $\beta=1/2$ cannot fully explain the observed impact of metaorders, nor its post-execution decay. 
In the following sections, we explore different routes to reconcile metaorder impact with long-term volatility.

\subsection{A generalized propagator model}

As we just discussed, the square-root price profile {\it during} the execution of the metaorder and the subsequent impact decay cannot be captured within the standard propagator model. Here we propose a (somewhat ad-hoc) extension of this model that allows one to decouple these two profiles. We will not attempt to fully justify such a proposal from first principles, but use the resulting equations as a convenient way to capture the known phenomenology of metaorder impact. 

Let us assume that once a metaorder has started, the market slowly adapts and, in the spirit of the LLOB model \cite{donier2015fullyconsistentminimalmodel, bouchaud2018trades}, progressively provides more liquidity to absorb the incoming flow. We represent this effect  as a {\it two-time} propagator, describing the impact of a child order occurring at time $t'$ after the start of the metaorder on the price at time $t > t'$: 
\begin{equation}\label{eq:new_propagator}
    G(t' \to t) = \frac{\theta(q)}{(\widetilde \varphi t' + n_0)^{1/2-\beta}} \left(\frac{\tau_0}{t-t' +\tau_0}\right)^{\beta}, \qquad (\beta < \frac12)
\end{equation}
where $\tau_0$ is the average time between two trades and $\widetilde \varphi \times t'$ is {\it number} of child orders executed since the start of the metaorder, which have eaten into the LLOB and therefore reveal more hidden liquidity. $n_0$ is the number of trades after which the metaorder is statistically detected by liquidity providers. 
The immediate impact of a child order is thus 
\begin{equation}
 G(t' \to t') =  \frac{\theta(q)}{(\widetilde \varphi t' + n_0)^{1/2 - \beta}}
\end{equation} 
which decreases with $t'$, as liquidity adapts. 

The impact of a metaorder of duration $s \gg \tau_0$ and $\widetilde \varphi s \gg n_0$ is now given by \footnote{Note that when $s \gg \tau_0$ but $\widetilde \varphi s \lesssim n_0$, impact behaves as in the standard propagator model as $t^{1 - \beta}$. If $\beta \approx 0.2$, such a behaviour is much less concave than a square-root, in agreement with the results of \cite{bucci2019crossover, maitrier2025double}.}
\begin{equation}\label{eq:prop_impact2}
    \mathcal{I}(t \leq s) \approx \theta(q) \int_0^t {\rm d}t' \, \frac{\widetilde \varphi^{1/2+\beta}}{t'^{1/2-\beta}}\left(\frac{\tau_0}{t-t'}\right)^{\beta}  = \mathcal{I}_1(q,\widetilde \varphi)\sqrt{t}, \quad \mathcal{I}_1(q,\widetilde \varphi):=\mathcal{B}_\beta \, \sqrt{\widetilde \varphi}  \theta(q) (\widetilde \varphi \tau_0)^\beta  ,
\end{equation}
where $\mathcal{B}_\beta=2\Gamma(1/2+\beta)\Gamma(1-\beta)/\sqrt{\pi}$. 

After the end of the metaorder, impact now decays as
\begin{equation}\label{eq:prop_decay2}
    \mathcal{I}(t > s)  = \mathcal{I}_1(q)  \sqrt{s}\left[\left(\frac{t}{s}\right)^{1-\beta} - \left(\frac{t}{s}-1\right)^{1-\beta}\right],
\end{equation}
which reproduces the empirical decay of metaorders if one chooses $\beta \approx 0.2$. With $\theta(q) \propto \sqrt{q}$, the peak impact then reads
\begin{equation}
    \mathcal{I}(Q) \propto (\widetilde \varphi \tau_0)^\beta \sqrt{Q},
\end{equation}
which still depends on the participation rate, but now with a smaller exponent $\beta=0.2$, more difficult to exclude empirically.

One could have hoped that the slower relaxation of impact in the post-execution regime would help recover a linear-in-$T$ behaviour of $\Sigma_T^2:=\mathbb{E}[\Delta_T^2]$. Unfortunately, one finds that the contribution of metaorder impact to $\Sigma_T^2$ scales as
\begin{equation}\label{eq:vol_long_range}
    \Sigma_T^2 \propto_{T \to \infty} 
    \begin{cases} 
        T^{1 - \gamma}, & \quad \gamma < 2\beta; \\
        T^{1 - 2\beta}, & \quad \gamma > 2\beta,
    \end{cases}
\end{equation}
which is again sub-diffusive whenever $\beta > 0$ and $\gamma = \mu - 1 > 0$.  In other words, impact-driven price diffusion is only possible when impact is permanent, i.e. $\beta=0$. This is in fact the assumption made by Sato \& Kanasawa in their latest paper \cite{sato2025exactlysolvablemodelsquareroot}. However, all empirical data known to us suggest that impact decays, at least over short to medium time scales \cite{moro2009market, brokmann2015slow, bucci2018slow, maitrier2025generatingrealisticmetaorderspublic}, which according to our calculation should lead to substantial price mean reversion on such time scales.

We now discuss a possible resolution of the diffusion ``paradox'', based on the autocorrelation of the sign of metaorders, which will be further validated in the next section. Two other proposals are discussed in Appendix A. 

\begin{figure}[H]
    \centering
    \includegraphics[width=0.7\linewidth]{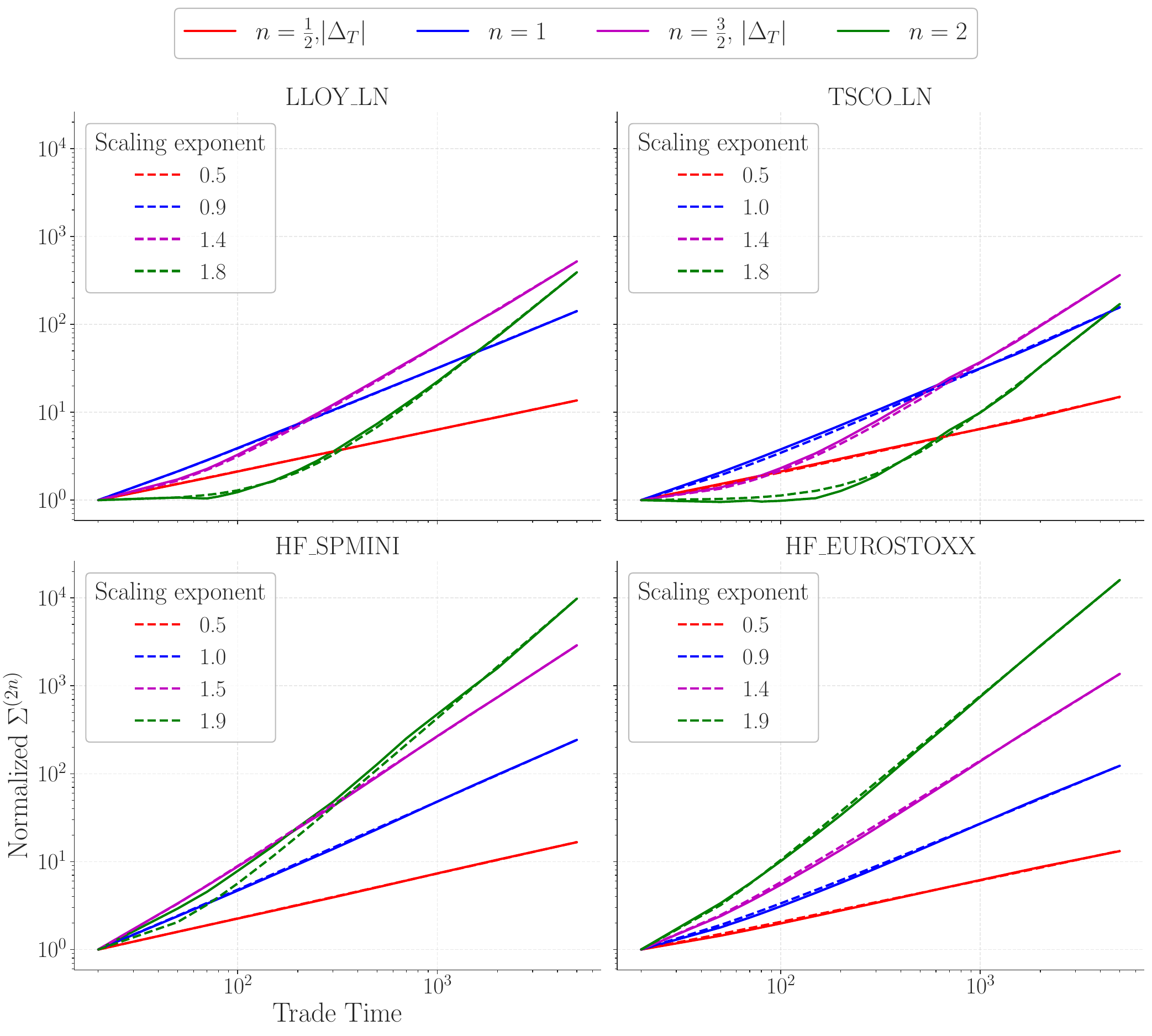}
    \caption{Scaling of the moments of price changes $|\Delta_T|^{2n}$ as a function of trade time $T$. We normalized the moment values such that all curves begin at $1$ for $T=1$. As shown, a sublinear behavior is found at short times for $n > 1$, likely attributable to price jumps that may dominate for short time scales when $n$ increases. Therefore, we fitted the data as $\Sigma^{(2n)} = a_0 + a_1 T^{\zeta_n} $ and present the values of $\zeta_n$ in the legend. We find in all cases $\zeta_n \approx n$, up to subleading corrections that we attribute to volume fluctuations.}
    \label{fig:scaling_moment_price}
\end{figure}

\subsection{Metaorder autocorrelations to the rescue}\label{subsec:diffusion_correlation_meta}

What happens if we assume, as in section \ref{sec:gamma_m}, that metaorders themselves are autocorrelated, with a new exponent $\gamma_\times$? Extending the calculation of $\mathbb{E}[\Delta_T^2]$ the case $\gamma > \beta - 1/2$ (always satisfied when $\beta < \frac12$), we find that these correlations contribute to volatility as 
\begin{equation}\label{eq:vol_long_range_od}
    \mathbb{E}[\Delta_T^2]_{o.d.} \propto \Gamma\, T^{2 - \gamma_\times - 2 \beta}.
\end{equation}
This leads to diffusion provided $\gamma_\times = 1 - 2 \beta$, which is, not surprisingly, the same condition as for the standard propagator model but with $\gamma$ replaced by $\gamma_\times$. Indeed, after coarse graining metaorders into effective single orders, we are back to the usual propagator model. 

For $\beta = 0.2$, this gives $\gamma_\times = 0.6$, which is close to $\gamma$ itself and close to its empirical value \cite{donier2015million} and G. Maitrier (unpublished). It is also compatible with the bound obtained in Eq. \eqref{eq:gamma_m}, which ensures that the autocorrelation of signs is dominated by the size distribution of metaorders, as postulated by \cite{lillo2005theory} and firmly established in \cite{sato2023inferring}.

The resulting value of volatility $\sigma^2$, defined as $\Sigma_T^2/T$, is given by
\begin{equation}
    \sigma^2 = C(\beta,\gamma_\times) \Gamma \nu^2 \tau_0^{\gamma_\times} \mathcal{I}_1^2(q, \varphi) \left(\overline{s^{\frac12 + \beta}}\right)^2 + \sigma^2_F,
\end{equation}
where $C(\beta,\gamma_\times)$ is a numerical coefficient found to be $\approx 5.6$ for $\gamma_\times=0.6$ and $\beta=0.2$, and $\sigma_F$ is the possible ``fundamental'' component of volatility (i.e. news driven).  

With $\theta(q) = \theta_0 \sqrt{q}$ and $\phi = \nu q \varphi \bar{s}$, we finally find 
\begin{equation}\label{eq:Y_od}
    \theta_0 = Y \frac{\sigma}{\sqrt{\phi}}, \qquad Y \propto \frac{\bar{n}^{\frac12 - \beta}}{\sqrt{C\Gamma}}\times \sqrt{1 - \frac{\sigma_F^2}{\sigma^2}}
\end{equation}
where we have assumed that $(\overline{s^{\frac12 + \beta}})^2 \propto \bar{s}^{1+2 \beta}$, which is justified in the present case since the $(1/2 + \beta)$-th moment of $s$ converges (whenever $\mu = 1 + \gamma > \frac12 + \beta$). We recall that $\bar{n} = \varphi \bar{s}$ is the average number of child orders per metaorder. 

This result is interesting since the peak impact is then precisely given by the standard square-root law, up to a \emph{weak} participation rate dependence when $\beta=0.2$:
\begin{equation}
    \mathcal{I}(Q) \propto \left({\widetilde \varphi \tau_0}\right)^\beta \sigma \sqrt{\frac{Q}{\phi}}.
\end{equation}
Note that we find naturally that impact is proportional to volatility, simply because volatility is due to impact!

The above calculation can be generalized to higher moments of $\Delta_T$. Assuming Wick-like factorisation of $\mathbb{E}[\varepsilon(t_1)\cdots \varepsilon(t_{2n})]$ as 
\begin{equation}
    \mathbb{E}[\varepsilon(t_1)\cdots \varepsilon(t_{2n})] \propto \underbrace{\prod_{i \neq j}|t_i - t_j|^{-\gamma_\times}}_{\text{$n$ pairs}},
\end{equation}
it is easy to show that the $2n$-th moment of $\Delta_T$ scales as:
\begin{equation}
    \Sigma_T^{(2n)} \propto T^{2n(1 - \beta) - n\gamma_\times} = T^n,
\end{equation}
as indeed observed empirically (up to subleading corrections that can also be rationalized within our framework), see Fig. \ref{fig:scaling_moment_price}. Note that we do {\it not} observe mutifractality (i.e. $\Sigma_T^{(2n)} \propto T^{\zeta_n}$ with $\zeta_n \neq n$) because we work in trade time and not in real time. As it is well known, multifractal effects come from intermittent fluctuations of the activity rate $\nu$, see e.g. \cite{muzy2000modelling, jusselin2018noarbitrageimpliespowerlawmarket}. An interesting extension of our model, which we leave for future work, would be to assume that $\nu$ itself has fractal properties.

The corrections to Eqs. \eqref{eq:vol_long_range}, \eqref{eq:vol_long_range_od} caused by volume fluctuations and a possible dependence of $\beta$ on $q$ are further discussed in Appendix A.

\section{Covariance between order flow imbalance and prices changes}\label{sec:SQLvsAgg}

\begin{figure}[H]
    \centering
    \includegraphics[width=0.6\linewidth]{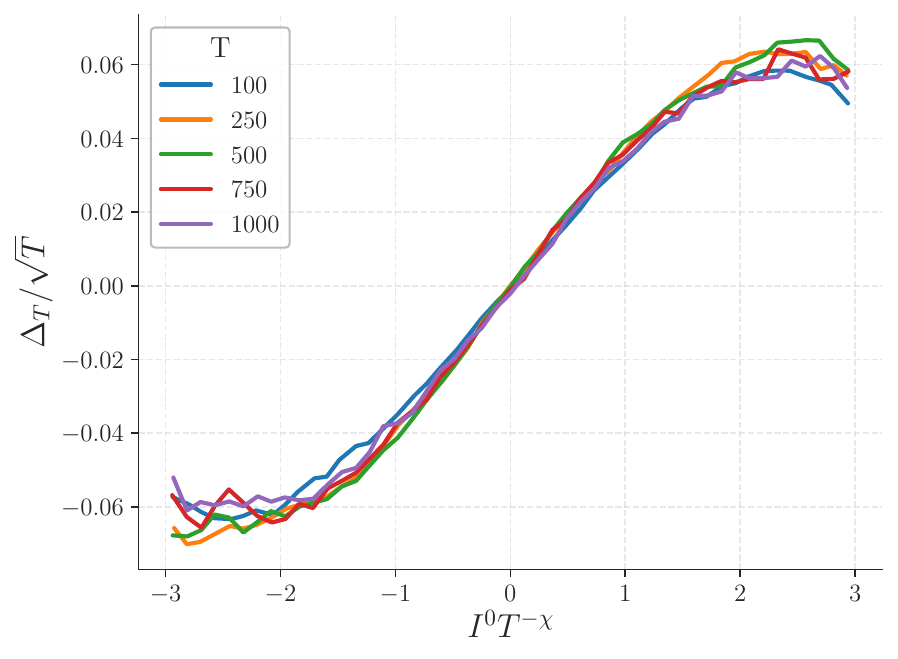}
    \caption{Aggregated impact versus sign imbalance for the EUROSTOXX. After appropriate rescaling, curves for different $T$ nicely collapse onto a single master curve. The y-axis rescaling reflects the diffusive nature of the price, while the x-axis rescaling captures trade sign correlations, as detailed in section \ref{sec:scaling_imbalances}. In this case, we find $\omega = \chi - \frac12  \approx 0.22$.
}
    \label{fig:Felix_Eurostoxx}
\end{figure}

An interesting quantity that can be computed within our model and easily measured empirically using the public tape of trades and prices is the ``aggregated'' impact $\mathbb{E}[\Delta | I^a]$, conditioned to a certain value of imbalance $I^a$, as studied in \cite{plerou2002quantifying,patzelt2018universal}. We know that such a quantity behaves very differently from the square-root law, and has non-trivial scaling properties as a function of $T$, see \cite{patzelt2018universal} for $a=0$ and $a=1$. For $a=0$, in particular, one finds that the initial slope of $\mathbb{E}[\Delta | I^0]$ as a function of $I^0$ scales like $T^{-\omega}$ with $\omega \approx 1/4$ \cite{patzelt2018universal, bouchaud2018trades}, a result we confirm in Fig. \ref{fig:Felix_Eurostoxx}.  

Note that if $I^a$ and $\Delta$ were Gaussian variables one could use the following general relation to predict that slope:
\begin{equation} \label{eq:cond_I}
    \mathbb{E}[\Delta | I^a] = \frac{\mathbb{E}[\Delta \cdot I^a]}{\Sigma_a^2} I^a,
\end{equation}
i.e. a linear aggregate impact for small imbalances, where $\Sigma_a^2$ was defined in section \ref{sec:volume_fluct}. However, in our model the Gaussian assumption does not hold since $I^a$ is a truncated Lévy variable (see section \ref{sec:volume_fluct}). Hence the exact calculation of $\mathbb{E}[\Delta | I^a]$ is much more intricate, and we restrict the following analysis to the covariance $\mathbb{E}[\Delta \cdot I^a]$, which we compare to empirical data below. Still, naively applying Eq. \eqref{eq:cond_I} will predict a scaling in $T^{-\omega}$, albeit within an uncontrolled approximation.

\subsection{Correlated metaorders and volume fluctuations}

In this subsection, we compute the covariance $\mathbb{E}[\Delta \cdot I^a]$ in the full blown model, with correlated metaorders and volume fluctuations. As above, we assume a $q$-dependent value of $\mu_q$ as given by Eq. \eqref{eq:muq} and further postulate a similar behaviour for $\beta_q$, i.e. 
\begin{equation}
 \beta_q = \beta_1 - \lambda' \ell,   
\end{equation} 
with $\ell=\log q$ (see also Appendix A). 

Now, there exists a value $q = q_c'$ such that $5/2 - \mu_{q_c'} = 1 - \beta_{q_c'}$. 
Then, using $\mathcal{I}_1(q,\varphi) \propto \sqrt{q}$, one gets, for the diagonal contribution
\begin{align}
    \mathbb{E}_q[\Delta_T \cdot I_T^a]_{d.} \propto \int_{0}^{q_c'} {\rm d}q \, \Xi(q) \, q^{a+\frac12} T^{5/2 - \mu_q} + \int_{q_c'}^\infty {\rm d}q \, \Xi(q) \, q^{a+\frac12} T^{1 - \beta_q} ,
\end{align}
and for the off-diagonal contribution
\begin{align}
    \mathbb{E}_q[\Delta_T \cdot I_T^a]_{o.d.} \propto \Gamma \, \mathbb{E}[q^a]  \int_{0}^\infty {\rm d}q \, \Xi(q) \, q^{\frac12} T^{2 - \gamma_\times - \beta_q} ,
\end{align}
Repeating the same calculations as in section \ref{sec:volume_fluct}, we now get:
\begin{align}
 \mathbb{E}_q[\Delta_T \cdot I_T^a]_{d.} &\propto  T^{5/2- \mu_m} \int_0^{\ell_c'} {\rm d}\ell \, \, e^{(\ell - m) (a + \frac12 - \lambda \log T) - (\ell - m)^2/2 \sigma_\ell^2} \nonumber \\
 &+ T^{1 - \beta_m} \int_{\ell_c'}^\infty {\rm d}\ell \, \, e^{(\ell - m) (a + \frac12 + \lambda' \log T) - (\ell - m)^2/2 \sigma_\ell^2}.
\end{align}
When $(\lambda, \lambda') \log T \ll 1$, the Gaussian integrals are dominated by the region around $\ell^\star = m + \sigma_\ell^2(a+1/2)$. So, schematically, when $\ell^\star < \ell_c' := \log q_c'$ the first integral dominates, while for $\ell^\star > \ell_c'$ the second integral dominates.
Hence, the dominant term scales as:
\begin{equation} \label{eq:scaling_IDelta_q}
     \mathbb{E}_q[\Delta_T \cdot I_T^a]_{d.} \propto e^{m (a + \frac12) + \frac12 \sigma_\ell^2 (a + \frac12)^2}  \begin{cases}
        & T^{5/2 - \widehat \mu(a)}, \qquad \widehat \mu(a) = \mu_m + (a + \frac12) \lambda \sigma_\ell^2 \qquad \qquad a < a_c' ;\\
        & T^{1 - \widehat \beta(a)} , \quad \qquad \widehat \beta(a) = \beta_m -  \left(a + \frac12\right)\lambda' \sigma_\ell^2  \qquad \quad a > a_c',
    \end{cases}
\end{equation}
where $a_c'$ is such that $\widehat \mu(a_c') = \mu_{q_c'}$. 

The power of $T$ coming from the diagonal contribution is thus predicted to {\it decrease} with $a$ when $a < a_c'$ and then to {\it increase} with $a$ for larger $a$. Hence we expect an interesting non-monotonic behaviour of the effective exponent as a function of $a$, i.e. with the relative weight given to child orders with large volumes. 

The off-diagonal contribution, on the other hand, gives:
\begin{equation} \label{eq:scaling_IDelta_q2}
     \mathbb{E}_q[\Delta_T \cdot I_T^a]_{o.d.} \propto \Gamma \,  e^{m (a + \frac12) + \frac12 \sigma_\ell^2 (a^2 + \frac14)} \, 
         T^{2 - \gamma_\times - \widehat \beta(0)}.
\end{equation}
Note that the coefficient in front of the power-law is $\Gamma e^{-\sigma_\ell^2 a/2}$ smaller than the one corresponding to the diagonal contribution. Other numerical prefactors may however contribute as well, that were neglected in the rough estimate of the above integrals. The final theoretical prediction is that $\mathbb{E}_q[\Delta_T \cdot I_T^a]$ is the sum of three power-law contributions: 
\begin{itemize}
    \item $T^{5/2 - \widehat \mu(a)}$, with an exponent equal to $1$ for $a=0$ and decreasing as $a$ increases, 
    \item $T^{1 - \widehat \beta(a)}$, with an exponent equal to $\approx 0.8$ for $a=0$ and increasing as $a$ increases, 
    \item $T^{2 - \gamma_\times - \widehat \beta(0)}$, with an exponent independent of $a$ and equal to $\approx 1.2$ for the default values of $\gamma_\times, \widehat \beta(0)$.
\end{itemize}    
In section \ref{sec:emp_cov} below, we will show that empirical data can be fitted as a power-law of $T$, with an effective exponent that indeed behaves non-monotonically with $a$, which suggests $(\lambda,\lambda')\sigma_\ell^2$ in the range $0.1$--$0.2$, also in line with the condition already obtained in section \ref{sec:volume_fluct}. 

Finally, note that the slope exponent $\omega$, naively predicted from Eq. \eqref{eq:cond_I} for $a=0$ is 
\begin{equation}
    \omega_{d.} = \frac12 + \widehat \mu(0) - \widetilde \mu(0) = \frac12 (1 + \lambda \sigma_\ell^2), \qquad  \omega_{o.d.} = 1 - \widetilde \mu(0) + \gamma_\times + \widehat \beta(0),
\end{equation}
depending on whether the diagonal or off-diagonal contribution dominates. Numerically, with $\lambda \sigma_\ell^2=1/8$ and $\gamma_\times + \widehat \beta(0) = 0.8$, one finds $\omega_{d.} \approx 0.56$ and $\omega_{o.d.} \approx 0.30$, close to the empirical value $1/4$ in the second case.

\subsection{The role of ``informed'' metaorders}\label{subsec:pref_informed}

Up to now, we assumed that there are no correlations between the sign of metaorders $\varepsilon$ and the ``fundamental'' component of price changes on time scale $T$, $\sigma_F \xi \sqrt{T}$, see Eq. \eqref{eq:price_dyn}. If we rather assume that $\mathbb{E}[\varepsilon \xi] = \rho/\sqrt{\nu T}$, where $\rho$ measures the average amount of information of individual metaorders,\footnote{For a detailed discussion of the scaling with $T$, see \cite{bouchaud2010impact}, section 16.1.3. The idea is that out of $N_T = \nu T$ metaorders of random signs, an excess fraction $\sim \sqrt{\nu T}$ is possibly informed. It is important to stress that, in the spirit of the Kyle model \cite{kyle1985continuous}, the fundamental component $\sigma_F \xi \sqrt{T}$ is not a mechanical consequence of impact.} we find an extra contribution to $\mathbb{E}[\Delta_T \cdot I_T^a]$ computed above, which reads:
\begin{align}\label{eq:pref_informed}
    \mathbb{E}[\Delta_T \cdot I_T^a] = \rho  \sqrt{\nu} \varphi q^{a} \sigma_F 
    \int_0^T {\rm d}u \int_0^\infty {\rm d} s \Psi(s) \left[\mathbb{I}(s > u) u + \mathbb{I}(s < u) s \right].
\end{align} 
Extending the calculation above, we now find that this covariance scales as $T^{2 - \mu}$ for $\mu < 1$ and as $T$ for $\mu > 1$, which is the case we focus on here. 

Hence, we find that such a fundamental contribution predicts a linear scaling of $\mathbb{E}[\Delta_T \cdot I_T^a]$ as a function of $T$, {\it independently} of $a$. Note that the naive prediction for the slope exponent $\omega$ (from Eq. \eqref{eq:cond_I}) is $\omega= 2 - \mu$ for $\mu \in [1,2]$.
 


\subsection{From covariances to correlations}\label{sec:correlation_coef}

Finally, we turn our attention to a natural description of the interplay between (generalized) volume imbalance and price changes, namely the  correlation coefficient
\begin{equation}\label{eq:corr}
    R_a(T) := \frac{\mathbb{E}[\Delta_T \cdot I_T^a]}{\Sigma_T \Sigma_a}, \qquad \Sigma_T:= \sqrt{\mathbb{E}[\Delta_T^2]}, \qquad \Sigma_a:=\sqrt{\mathbb{E}[I_T^{a2}]}
\end{equation}
In order to simplify the discussion, we assume that the crossovers between the two regimes for $\Sigma_a^2$ (Eq. \eqref{eq:scaling_Ia_2}) and for $\mathbb{E}[\Delta_T \cdot I_T^a]$ (Eq. \eqref{eq:scaling_IDelta_q}) occur for the same value of $a = a_c = a_c'$. Although this is not precisely true, the following conclusions will be qualitatively correct.

In the case where the sign of metaorders and the fundamental component of price changes are independent ($\rho=0$) and the off-diagonal contribution can be neglected ($\Gamma = 0$), we find that for $T \gg 1$: 
\begin{equation} \label{eq:scaling_corr}
    R_a(T)_{d.} \propto e^{\frac{\sigma_\ell^2}{2}a(1-a)} \times \begin{cases}
        & T^{(1- \mu_m  - \lambda \sigma_\ell^2)/2}, \quad \quad  a < a_c; \\
        & T^{- \widehat \beta(a)}, \qquad \qquad  \quad a > a_c.
    \end{cases}
\end{equation}
A more refined analysis would be needed for small values of $T$, but from such an analysis we conclude that $R_a(T)$ should be decreasing with $T$ for small values of $a$ (since $\mu_m \approx 3/2$) and saturating for large values of $a$ (since $\widehat \beta(a > a_c') = 0$). Furthermore, note the non-monotonic behaviour (in $a$) of the prefactor in Eq. \eqref{eq:scaling_corr}. 

If we now consider the contribution coming from the correlation between metaorders, we get:
\begin{equation} \label{eq:scaling_corr2}
    R_a(T)_{o.d.} \propto  e^{-\frac{\sigma_\ell^2 a^2}{2}} \times 
         T^{\mu_m/2 + a \lambda \sigma_\ell^2 - \gamma_\times - \widehat \beta(0)}.
\end{equation}
For $a \to 0$, the power of $T$ is very close to zero for our default choice of parameters. For higher values of $a$, the exponent becomes positive and therefore this off-diagonal contribution adds an increasing function of $T$.\footnote{In the asymptotic $T \to \infty$ limit, one should take into account the fact that $\Sigma_a$ itself becomes dominated by the off-diagonal contribution, see the discussion around Eq. \eqref{eq:sigma_a_od}. Hence the correlation does saturate when $T \to \infty$, as it should be.}

If we assume instead that $R_a(T)$ is dominated by informed metaorders, we obtain, in the regime where $\mu_q > 1 + \beta_q$, $\forall q$,
\begin{equation} \label{eq:scaling_corrF}
    R_a(T)_F \propto \rho e^{-\frac{\sigma_\ell^2 a^2}{2}} \times \begin{cases}
        & T^{\mu_m/2 + a \lambda \sigma_\ell^2 - 1}, \quad \quad  a < a_c; \\
        & T^{0}, \qquad \qquad \, \quad a > a_c.
    \end{cases}
\end{equation}
The scaling with $T$ indicates that $R_a(T)$ should decrease with $T$ for small values of $a$, and become independent of $T$ for large values of $a$. Note that if volatility is mostly due to fundamentals and not due to impact, one finds that $R_a(T)$ is directly proportional to $\rho$.\footnote{It may be realistic to assume that well informed metaorders are larger, and therefore that $\rho \propto q^\psi$ where $\psi \geq 0$. In this case, the scaling with $a$ reads $R_a(T)_F \propto e^{{\sigma_\ell^2} a (2 \psi -a)/2}$, which reaches a maximum for $a^\star=\psi$. }

Hence, we see that the three possible contributions to $R_a(T)$ have different monotonicity properties as functions of $T$, suggesting that different shapes of $R_a(T)$ might be observed empirically. In the following, we will confirm that this is indeed the case. The behaviour of $R_a(T)$ for a given $T$ as a function of $a$ is simpler to describe. One finds that for $a < a_c$
\begin{equation} \label{eq:R_a_vs_a}
    R_a(T) =  e^{-\frac{\sigma_\ell^2 a^2}{2}} \left(A(T) e^{\frac{\sigma_\ell^2 a}{2}} + B(T)  e^{\lambda \sigma_\ell^2 a\log T }\right),
\end{equation}
where $A,B$ are functions of $T$, the second contribution $B(T)$ coming from the $a$ dependence of the exponents. Hence we expect a behaviour of $R_a(T)$ that always increases for small $a$, independently of the dominant contribution (diagonal, off-diagonal or fundamental), but with a slope that increases with $T$ in the last two cases. These predictions will be tested against empirical data in the next section.

\section{Covariances \& Correlations: Empirical data}\label{sec:emp_cov}

The theoretical analysis laid out in the previous sections makes several non-trivial predictions:
\begin{enumerate}
    \item  Provided the main source of price moves is the {\it average} impact of random metaorders, the covariance $\EX[\Delta_T \cdot I_T^a]$ behaves as a power-law of $T$ with an effective exponent that is non-monotonic in $a$, reaching a minimum for some value of $a$, see Eq. \eqref{eq:scaling_IDelta_q}. If instead volatility is dominated by the random component of impact (Eq. \eqref{eq:price_dyn}) with a ``fundamental'' component $\sigma_F \xi$, a linear behaviour in $T$ independent of $a$ should be observed;
    \item The correlation coefficient $R_a(T)$ contains an off-diagonal contribution growing with $T$ and two contributions (diagonal and fundamental) decaying with $T$ before saturating. Depending on the relative amplitude of these contributions, different shapes can be expected;
    \item For a given $T$, the correlation coefficient $R_a(T)$ is predicted to be a humped shape function of $a$.
\end{enumerate}
We will show below that, quite remarkably, all these predictions are in qualitative agreement with empirical data. This will enable us to estimate the parameters $\lambda$ and $\lambda'$. We will also see a marked difference between large tick assets like EUROSTOXX and smaller tick assets (LLOY, TSCO and SPMINI).

\begin{figure}[H]
    \centering
    \includegraphics[width=0.7\linewidth]{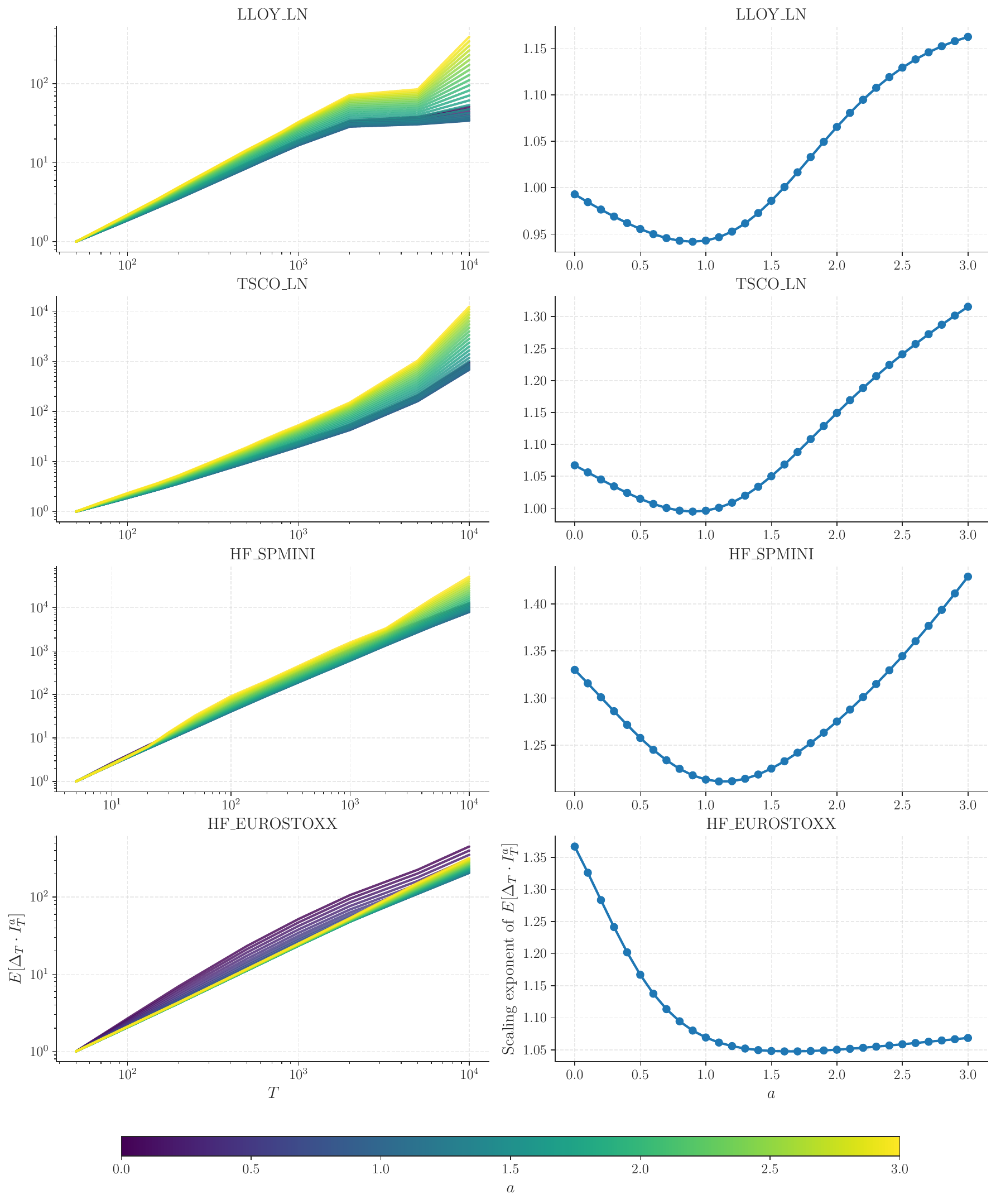}
    \caption{Covariance $(\Delta_T, I^a_T)$ as a function of $(T, a)$ for the four considered assets. \textbf{Left:} Log-log plot of $\EX[\Delta_T \cdot I_T^a]$ vs. $T$ for different values of $a$. \textbf{Right:} Scaling exponents as a function of $a$, obtained by fitting the initial regime ($T < 10^3$).} 
    \label{fig:cov_vs_T_and_a}
\end{figure}

\subsection{Power-law behaviour of the covariance}

We first investigate point 1. above. As shown in Fig. \ref{fig:cov_vs_T_and_a} (left), a power-law behaviour of $\EX[\Delta_T \cdot I_T^a]$ as a function of $T$ is approximately verified for all $a$, albeit with some amount of convexity (for TSCO) or concavity (for LLOY or EUROSTOXX). When plotted as a function of $a$, the effective exponent of $T$ displays the predicted non-monotonic behaviour, see Fig. \ref{fig:cov_vs_T_and_a} (right), reaching a minimum for $a \approx 1$ for LLOY and TSCO, $a \approx 1.1$ for SPMINI and $a \approx 1.5$ for EUROSTOXX. The left slope is predicted to be equal to $- \lambda \sigma^2_\ell$ and the right slope to $+ \lambda' \sigma^2_\ell$. For LLOY, TSCO and SPMINI we thus find $\lambda \sigma^2_\ell \approx 0.1$ (not far from the estimate derived from Fig. \ref{fig:empiricalIa}) and $\lambda' \sigma^2_\ell \approx 0.15 $--$0.2$. For EUROSTOXX, we estimate $\lambda \sigma^2_\ell \approx 0.5$, a factor 2 larger than from the behaviour of $\Sigma_a$ (Fig. \ref{fig:empiricalIa}), but a very small (but positive) value for $\lambda' \sigma^2_\ell$. 

The value of the effective exponent is in the range $0.95$ -- $1.35$, as expected since the diagonal contributions give an exponent slightly below $1$ (Eq. \eqref{eq:scaling_IDelta_q}) and the off-diagonal contribution yield an exponent $\approx 1.2$ for the default values of $\gamma_\times$ and $\widehat \beta(0)$ (see Eq. \eqref{eq:scaling_IDelta_q2}).

Note that, as mentioned above, a volatility model based on fundamentals only, leads to a linear behaviour  $\EX[\Delta_T \cdot I_T^a] \propto T$,  independently of the value $a$, clearly at odds with Fig. \ref{fig:cov_vs_T_and_a} (right). For the EUROSTOXX, such a linear regime can perhaps be observed for $a \gtrsim 1.5$, but also compatible with Eq. \eqref{eq:scaling_IDelta_q} if $\lambda'$ is small. 

\subsection{Correlation vs. $T$ and $a$}

\begin{figure}[H]
    \centering
    \includegraphics[width=0.7\linewidth]{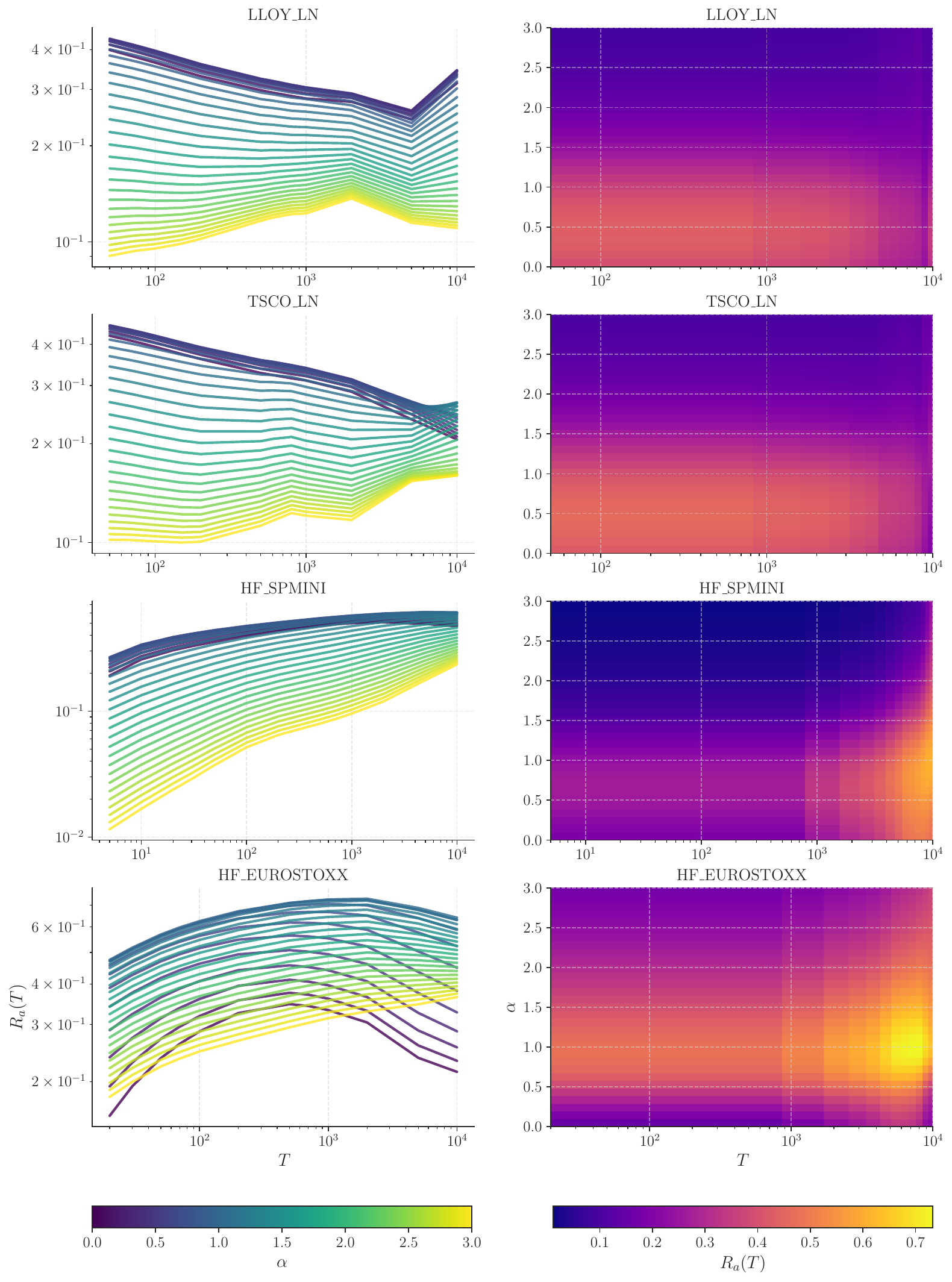}
    \caption{Analysis of the correlation function $R_a(T)$ for LLOY, TSCO, SPMINI and EUROSTOXX. \textbf{Left column:} Evolution of the correlation for different values of $a$, showing the non monotonic behavior \textbf{Right column:}  Heatmap illustrating the distribution of correlation values within the $(a,T)$ space, indicating that the correlation reaches its peaks for $a \approx 0.5 - 1$, regardless of the T values.  }
    \label{fig:corr_vs-T_and_a}
\end{figure}

Turning to point 2., Fig. \ref{fig:corr_vs-T_and_a} shows the correlation coefficient $R_a(T)$ vs. $T$ for different $a$ in two different representations: standard plot and heatmap. A first immediate observation is that these correlations are $O(1)$ for all $T$ values, and peak around $0.45$ for stocks and $0.7$ for futures. This means that order flow and returns are indeed strongly correlated, as was emphasized many times (see e.g. \cite{evans2002order, chordia2004order, hopman2007supply, bouchaud2009markets, patzelt2018universal}). 

We observe that for LLOY and TSCO, $R_a(T)$ is a mildly decreasing function of $T$ for small $a$, which becomes mildly increasing for larger $a$, as expected from the discussion in section \ref{sec:correlation_coef}, assuming that the diagonal contribution dominates for small $a$ and the off-diagonal contribution kicks in for larger $a$, or larger values of $T$ (as indeed suggested by the two upper plots in Fig. \ref{fig:corr_vs-T_and_a}), where an upturn of $R_a(T)$ is observed at large $T$.

For EUROSTOXX and SPMINI, we observe a non-monotonic behaviour of $R_a(T)$ vs. $T$ for small $a$, with a maximum reached for rather large values of $T$. This suggests that the non-diagonal contribution is dominant for small $a$, with an exponent that is already positive for $a=0$, i.e. values of $\mu_m$ and $\widehat \beta(0)$ larger than the default values quoted throughout the paper to be compatible with our admittedly rough theoretical analysis. The qualititative behaviour of both futures contracts thus appears to be quite different from that of stocks. Note in particular that the level of the correlation is markedly higher for EUROSTOXX, reaching a maximum value $\approx 0.75$, compared to $\approx 0.45$ for stocks (see also Fig. \ref{fig:corr_vs-T_and_a}) and $\approx 0.6$ for SPMINI.

Finally, note that the saturation regime where $R_a(T)$ should become independent of $T$, as predicted by either the ``diagonal'' hypothesis (\eqref{eq:scaling_corr}) or the ``fundamental'' hypothesis (Eq. \eqref{eq:scaling_corrF}), is hardly observable in the data, at least up to $10^4$ trades (roughly one trading day). This observation appears to confirm the predominant role of {\it impact} of mostly uninformed (but correlated) metaorders in the genesis of volatility \cite{bouchaud2018trades}.  

\subsection{Non-monotonic behaviour of $R_a$ vs. $a$}

\begin{figure}[H]
    \centering
    \includegraphics[width=0.6\linewidth]{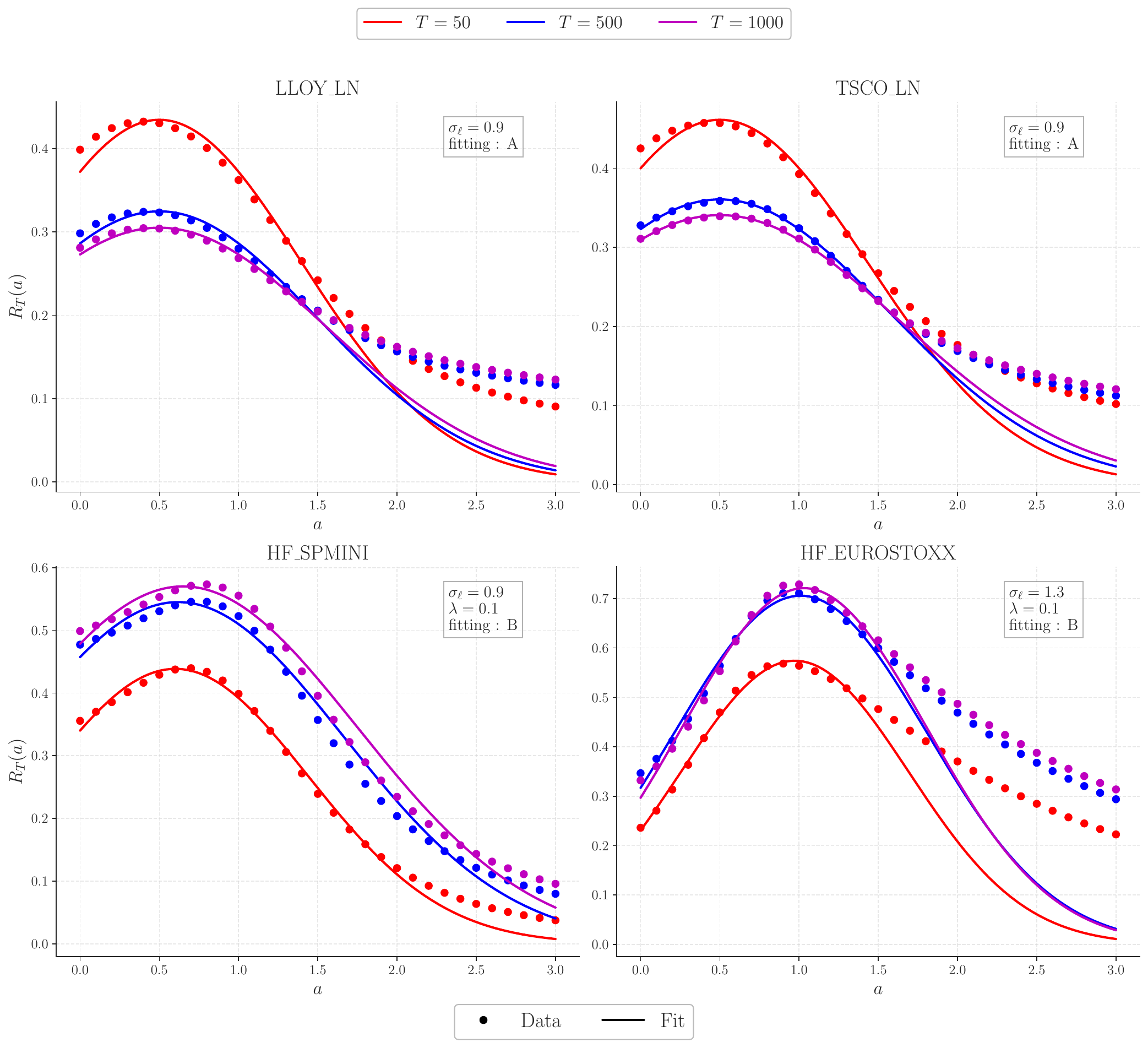}
    \caption{Fit of the correlation function $R_a(T)$, for several $T$. Since Eq.~\eqref{eq:R_a_vs_a} holds only for $ a < a_c $, we restrict the fit to $a < 1.5$. The empirical estimates of $ \sigma_\ell$ obtained with this fit turn out to be surprisingly close to the ones obtained in Fig. \ref{fig:lognormal_child_all}.}
    \label{fig:corr_with_fit}
\end{figure}

Finally, for point 3., Fig. \ref{fig:corr_with_fit} shows that the correlation between price returns and generalized order imbalance is non-monotonic as a function of $a$, reaching a maximum for $a^\star \approx 0.5$ for LLOY, TSCO and SPMINI, and $a^\star \approx 1$ for EUROSTOXX. 

Such a non-monotonic behaviour is predicted by our theoretical analysis. Interestingly, when the diagonal contribution to $R_a$ dominates, we expect that the maximum correlation is reached precisely for $a^\star=1/2$, with a peak amplitude that decreases with $T$, see Eq. \eqref{eq:scaling_corr}. The data for the two stocks is therefore compatible with the fact that in the small $a$ regime, $R_a(T)_{d.}$ is a decreasing function of $T$. In this regime, one can also predict that 
\begin{equation}\label{eq:R_ratio}
    \frac{R_{a=\frac12}(T)}{R_{a=0}(T)} = e^{\sigma_\ell^2/8},
\end{equation}
to be compared with the data for which this ratio is $\approx 1.1$. The inferred value of $\sigma^2_\ell$ is thus around $1$, comparable to the direct estimate of $\sigma^2_\ell$ from the variance of log-volumes, see section \ref{sec:volume_emp}. The full fit of $R_a(T)$ vs. $a$ neglecting that the $B$ term in Eq. \eqref{eq:scaling_corr} is given in Fig. \ref{fig:corr_with_fit}.

For the EUROSTOXX, on the other hand, the maximum is reached for larger values of $a$, and the ratio defined in \eqref{eq:R_ratio} is much larger (3 -- 5), suggesting that the $B(T)$ term in Eq. \eqref{eq:R_a_vs_a} is now dominant, as demonstrated by a fit to the data, see Fig. \ref{fig:corr_with_fit}. This is consistent with our remark above -- that the off-diagonal term dominates $R_a(T)$ for small $T$. In this scenario, the initial positive slope of $R_a(T)$ vs. $a$ should increase with $T$, in agreement with data. Neglecting $A(T)$ in Eq. \eqref{eq:scaling_corr} and setting $a^\star = 1$, we now obtain
\begin{equation}\label{eq:R_ratio2}
    \frac{R_{a=1}(T)}{R_{a=0}(T)} = e^{\sigma_\ell^2 (\lambda \log T - \frac12)},
\end{equation}
which can indeed become large: if we take $\sigma_\ell^2=2$ and $\lambda \sigma_\ell^2 = 1/2$, as suggested above, one finds that for $T=100$ the ratio above is $\approx 3.7$. For $\lambda \sigma_\ell^2 = 1/4$, that ratio is $\approx 1.1$.

\subsection{Empirical covariances \& correlations: summary}

All empirical data appear to confirm the qualitative validity of our predictions, some of them being rather non-trivial. One of the main assumptions of our model is that the exponents describing the autocorrelation of child orders ($\mu_q$) and the decay of their impact ($\beta_q$) depend on the size $q$ of these child orders. This, in turn, leads to a non-monotonic dependence of the scaling of the covariance $\EX[\Delta_T \cdot I_T^a]$ with $a$,\footnote{Such a non-monotonic behaviour cannot be explained simply from the power-tail distribution of volumes $q$, see G. Maitrier et al., in preparation.} and of the correlation $R_a(T)$ with both $T$ and $a$. Although some of our predictions fail to explain the data quantitatively, we are tempted to ascribe these discrepancies on the bluntness of our approximations, which, we argue, correctly capture the mechanisms at play. 

Perhaps the two most important conclusions of this section, beyond the success of our model in capturing the main trends of the covariance data, are:
\begin{itemize}
    \item Stocks and futures seem to differ quantitatively when it comes to the {\it correlation} between order imbalance and price changes. In particular, the correlation between the two is stronger for futures contract, and is reached for longer time intervals $T$ and with more weight given on large child order volumes (i.e. $a^\star=1$ instead of $a^\star=1/2$). This suggests a stronger role of metaorder correlations for futures than for stocks.
    \item The hypothesis according to which most of the volatility comes from fundamental information is hard to reconcile with the data. First, as discussed in section \ref{subsec:impactfluctuation}, this assumption does not enable one to rationalize why the square-root law, which applies to all metaorders (informed or not \cite{gomes2015market, bucci2018slow, sato2024does}) is proportional to volatility. Second, it predicts that the covariance of price returns and volume imbalances is proportional to $T$ {\it independently of } $a$, at odds with empirical data. As argued in \cite{bouchaud2006random, taranto2018linear, bouchaud2018trades}, the most plausible hypothesis is that volatility stems from trading alone. In other words, {\it the excess volatility puzzle has a microstructural origin}. 
\end{itemize}

\section{Conclusion}

The aim of this paper was to reconcile several apparently contradictory observations: is a square-root law of metaorder impact that decays with time compatible with the random-walk nature of prices and the linear impact of order imbalances? Can one entirely explain the volatility of prices as resulting from a ``soup'' of indistinguishable, randomly intertwined and uninformed metaorders? 

In order to answer these questions, we have introduced a new theoretical framework to describe metaorders with different signs, sizes and durations, possibly correlated between themselves, which {\it all} impact prices as a square-root of volume (which we assume as an input) but with a subsequent time decay characterized by an exponent $\beta \neq \frac12$, i.e. different from the one suggested by the classic propagator model \cite{bouchaud2003fluctuations, bouchaud2018trades} or the LLOB model \cite{donier2015fullyconsistentminimalmodel}. We proposed a generalized propagator model to account for such a feature. 

We then established that the power-law tailed distribution of metaorder durations is not sufficient to counteract impact decay that leads to price sub-diffusion. Rather, as in the original propagator model, price diffusion is ensured by the long memory of cross-correlations {\it between} metaorders, which is indeed present in data. In fact, we conjecture that the intra- and cross-correlations between child orders decay roughly in the same manner, a feature that may be crucial for explaining the success of the construction of synthetic metaorders from public data \cite{maitrier2025generatingrealisticmetaorderspublic}. Such a prediction is empirically falsifiable, and we look forward to more work in that direction using, e.g., the Tokyo Stock Exchange data of \cite{sato2023inferring, durin2023squarerootlawsmarket}.

The existence of correlations {\it between} metaorders is therefore a crucial ingredient to recover price diffusion. The old debate between order splitting and ``herding'' that seemed to have been closed by several papers in favor of splitting \cite{toth2015equity, sato2023inferring}, is perhaps not so clear-cut. Such correlations could be due either to the fact that many participants use the same trading signals, or that copy-cat metaorders/unwinding market maker inventories correlate to past order flow, or else that some traders successfully predict the future behaviour of other participants. Note that within our story any predictive alpha signal manifests itself through autocorrelated metaorders (see also the discussion in  \cite{bouchaud2018trades}, chapter 20). 

In view of the strongly fluctuating volumes $q$ of child orders, one quickly realizes that one needs to account for heterogeneity in the distribution of metaorder durations, and in the resulting decay of their impact, which we parametrized by two $q$-dependent exponents, $\mu_q$ and $\beta_q$, assumed to depend linearly on $\ell = \log q$. This feature allowed us to account semi-quantitatively for the way the moments of generalized volume imbalance scale with time $T$, and more importantly how the correlation between price changes and generalized volume imbalance scales with $T$. We predicted that the corresponding power-law should depend in a non-monotonic fashion on the parameter $a$ that allows one to put the same weight on all child orders ($a=0$) or overweight large orders ($a$ large), a behaviour clearly borne out by empirical data, see Fig. \ref{fig:cov_vs_T_and_a}. We also predicted that the correlation between price changes and volume imbalances should display a maximum as a function of $a$ for fixed $T$, which again matches observations, see Fig. \ref{fig:corr_with_fit}, with fitting parameters fixed with previously determined values. We found that stocks and futures appear to differ quite markedly in terms of these metrics, which could provide an interesting new way to characterize price formation mechanisms in different markets. 

Such noteworthy agreement between theory and data suggests that our framework correctly captures the basic mechanism at the heart of price formation, namely the average impact of metaorders. We claim that our results strongly support the ``Order-Driven'' theory of markets, according to which it is the mechanical impact of trades, independently of any notion of ``fundamental information'', that generates volatility in financial markets, a picture advocated in \cite{bouchaud2006random, hopman2007supply, bouchaud2018trades, bouchaud2022inelastic} and, in a different context, in \cite{gabaix2021search}. In particular, the Efficient Market Hypothesis, which posits that volatility is mostly due to the variation of ``fundamental value'', cannot easily explain our results concerning the correlations between price changes and order flow imbalance. 

All these ideas are quantitatively confirmed  using a numerical simulation of our model in a follow up of this work \cite{maitrier2025subtle}. Such a numerical implementation will allow us in to (i) generate realistic synthetic datasets that capture the subtle interplay between order flows and prices, and (ii) validate on such synthetic data the method of metaorder reconstruction from the public tape recently proposed in \cite{maitrier2025generatingrealisticmetaorderspublic}.

\subsection*{Note added} After submission of this paper, we received the very interesting preprint of Muhle-Karbe et al., \emph{A unified theory of order flow, market impact, and volatility} (January 30, 2026). The precise relation between our approach and theirs certainly deserves more investigations. We thank M. Rosenbaum for sharing his paper with us. 

\subsection*{Acknowledgments}

We wish to thank J. D. Farmer, J. Bonart, N. Hey, K. Kanazawa, J. Kurth, C. A. Lehalle, F. Lillo, G. Loeper, F. Patzelt, J. Ridgeway, M. Rosenbaum, Y. Sato \& B. T\'oth for many enlightening conversations on these topics. This research was conducted within the Econophysics \& Complex Systems Research Chair, under the aegis of the Fondation du Risque, the Fondation de l'\'Ecole Polytechnique and Capital Fund Management.

\subsection*{Disclosure of interest \& Funding}
The authors declare no conflicts of interest. No specific funding was received.

\printbibliography

\newpage 

\section*{Appendix A}
\setcounter{equation}{0}
\renewcommand{\theequation}{A\arabic{equation}}

In the main text, we argued that sign correlations between metaorders offered a way out of the diffusivity paradox. In this Appendix, we consider two other possible scenarios: one where diffusion comes from large uncorrelated metaorders with permanent impact; the second where impact {\it fluctuations} play a major role. We conclude this appendix by comparing the merits of these alternative scenarios.  

\subsection*{The role of volume fluctuations}\label{sec:volume_sqrt}

Another possibility is to take seriously the fact that the size $q$ of child orders is fluctuating and correlated with the duration $s$ of metaorders. As we have shown in section \ref{sec:volume_fluct}, a dependence of the exponent $\mu_q$ on $q$ explains how different {\it moments} of the volume imbalance depend on $T$, see Eqs. \eqref{eq:scaling_Ia_2}, \eqref{eq:scaling_Ia_n}.

Now, similarly to $\mu_q$, we might assume that the impact decay exponent $\beta$ becomes $q$-dependent: $\beta_q = \beta_1 - \lambda' \log q$.\footnote{A sufficient condition for price diffusivity in the standard propagator model is $\beta_1 = 1 - \mu_1/2$ and $\lambda' = \lambda/2$, but we will not impose such constraints below and leave $\lambda$ and $\lambda'$ free.} One then observes that price impact becomes permanent for large child orders, with $q > q_0$ such that $\beta_q \leq 0$. When metaorder signs are independent, the non-vanishing contribution to volatility then reads:
\begin{align}
   \mathbb{E}[\Delta_T^2]_{q > q_0} = \nu \int_{q_0}^{\infty} {\rm d}q\, \Xi(q) \mathcal{I}_1^2(q,\varphi) \int_0^T{\rm d}u\,  \int_0^{u} {\rm d}s\, \Psi_q(s) s \approx_{T \to \infty} \nu T  
   \int_{q_0}^{\infty} {\rm d}q\, \Xi(q) \mathcal{I}_1^2(q,\varphi) \bar{s}_q.
\end{align}
Taking $\theta(q)=\theta_0 \sqrt{q}$ one gets $\mathcal{I}_1(q > q_0,\varphi)=2\theta_0 \sqrt{\varphi q}$, so the contribution to long-term price volatility is given by
\begin{equation}
    \sigma^2 = \frac{\mathbb{E}[\Delta_T^2]_{q > q_0}}{T} \propto \theta_0^2 \phi_{0}, 
\end{equation}
where $\phi_{0}$ is the average volume flow of the market, restricted to ``large'' child orders $q > q_0$. Interestingly, this relation can be read backwards as
\begin{equation}
    \theta_0 \propto \frac{\sigma}{\sqrt{\phi_{0}}},
\end{equation}
allowing one to recover the full square-root impact law from the expression of $\mathcal{I}_1$ above:
\begin{equation}
    \mathcal{I}(Q|q,\widetilde \varphi) = Y(q,\widetilde \varphi) \sigma \sqrt{\frac{Q}{\phi}},\qquad Y(q,\widetilde \varphi) \propto (\widetilde \varphi \tau_0)^{\beta_q} \,  \sqrt{\frac{\phi}{\phi_0}}.
\end{equation}

We see that this result suggests a weak dependence of the prefactor of the square-root law in $q$ and $\widetilde \varphi$, which disappears for large enough $q > q_0$, since $\beta_q \to 0$ in that case. However, in that case $Y$ would be substantially larger than empirically observed. Besides, since child orders of size $< q_0$ (which are the most numerous) only impact prices temporarily, this scenario would again lead to strong visible mean-reversion in prices not observed in data -- see e.g. Fig. \ref{fig:scaling_moment_price} for $n=1$.

It is, however, important to discuss how volume fluctuations might affect the results Eqs. \eqref{eq:vol_long_range}, \eqref{eq:vol_long_range_od} above, induced by the correlation between different metaorders. It is plain to extend the calculations of section 3.3 to get

\begin{equation}\label{eq:vol_long_range2_od}
    \mathbb{E}[\Delta_T^2]_{o.d.} \propto \Gamma e^{m + \frac{\sigma_\ell^2}{4}} \, T^{2 - \gamma_\times - 2 \beta_m + \lambda' \sigma_\ell^2}, 
\end{equation}
whereas a similar calculation for the diagonal contribution gives
\begin{equation}\label{eq:vol_long_range2}
    \mathbb{E}[\Delta_T^2]_{d.} \propto e^{m + \frac{\sigma_\ell^2}{2}} \, T^{1 - 2 \beta_m + 2 \lambda' \sigma_\ell^2}. 
\end{equation}
The ratio of the prefactors is now only $e^{\sigma_\ell^2/4}/\Gamma$ in favor of the diagonal term. This means that the off-diagonal contribution, with a larger power of $T$, becomes dominant beyond a reasonable small value of $T$ when $\sigma^2_\ell = 1$ and $\Gamma=O(1)$. We will therefore choose in the following $\gamma_\times=0.6$ and $\widehat \beta(0):= \beta_m - \frac12 \lambda'\sigma_\ell^2 = 0.2$, such that the off-diagonal contribution is exactly diffusive. 

\subsection*{The role of impact fluctuations}\label{subsec:impactfluctuation}

Up to now, we have assumed {\it deterministic} impact and neglected the role of price changes induced by ``news'' or other order book events that are not related to trades, with the ambition of recovering all the price volatility from the impact of metaorders. However,  it is clear that:
\begin{itemize}
    \item (i) news events do obviously exist (see e.g. \cite{aubrun2025identifying} for a recent discussion) and should indeed contribute to volatility. In fact, Efficient Market Theory predicts that the {\it only} contribution to volatility comes from news!
    \item (ii) the impact of a given metaorder has no reason to be deterministic: it should depend on specific time-dependent market conditions and thus include a random component. 
\end{itemize} 
Such a random component was in fact indirectly observed by Bucci et al. \cite{bucci_impact_vol2019}, where it was found that the effect of a single metaorder on price changes reads 
\begin{equation}\label{eq:price_dyn}
   \Delta_T =  p_T - p_0 \approx \varepsilon \mathcal{I}(Q) \left[1 + z \eta \right],
\end{equation}
where $\eta$ is a zero mean, unit variance, independent random variable, and $z$ a coefficient measuring the relative fluctuations of impact, found to be around $3$ in \cite{bucci_impact_vol2019}.\footnote{Note that there is an error in that paper, where $z$, called $a$ there, was reported to be around $0.1$.} 

Let us postulate that, while the average impact decays to zero with time as per the propagator model, the random component $z\eta$ does not, or at least not completely. This assumption does not violate any known stylized facts about the {\it average} decay of impact. Following these ideas, we expect price changes $\Delta_T$ to include extra terms that read 
\begin{equation}
    \Delta_{T,1} = z_\infty \theta_0 \sqrt{q \varphi} \int_0^T {\rm d}N_t \,\eps(t) \left[\mathbb{I}(t+s > T) \sqrt{T-t}  + \mathbb{I}(t+s < T) \sqrt{s}  \right] \eta_t + \sigma_F \xi \sqrt{T},
\end{equation}
where $z_\infty \leq z$ accounts for a possible time decay of the random component of impact, and the last contribution captures fundamental ``news'', with a volatility $\sigma_F$. ($\xi$ is another zero mean, unit variance, independent random variable).

This gives rise the following contribution to volatility:\footnote{Note that since $\mathbb{E}[\varepsilon \eta]$ is assumed to be zero, there is no particular role for metaorder correlations in this scenario. One could however wonder how the result given in Eq. \eqref{eq:vol_long_range} might change if we assume ``informed metaorders'', i.e. some correlations between the sign of the metaorder $\varepsilon$ and the subsequent fondamental price change $\xi$, see section \ref{subsec:pref_informed}. The result is a contribution to $\Sigma_T^2$ proportional to $\rho T^{1 - \beta}$, which is subdominant at large $T$.}
\begin{equation}
    \Sigma_T^2 = \mathbb{E}[\Delta_T^2]_{\eta, \xi} = z_\infty^2 \theta_0^2 q \varphi \nu \int_0^T {\rm d}u \int_0^\infty {\rm d}s \Psi(s)
    \left[\mathbb{I}(s > u) u + \mathbb{I}(s < u) {s}  \right] + \sigma_F^2 T,
\end{equation}
whence a long-term volatility given by
\begin{equation} 
    \sigma^2 = \sigma_F^2 + z_\infty^2 \theta_0^2 q \varphi \nu \bar{s}  \equiv \sigma_F^2 + z_\infty^2 \theta_0^2 \phi,
\end{equation}
where $\phi = q \varphi \nu \bar{s}$ is, again, the average volume flow in the market.

This expression is quite interesting: inserting $\theta_0 = Y \sigma/\sqrt{\phi}$, we get, provided $Y z_\infty < 1$:\footnote{Note that the same expression would hold with $z_\infty=1$ and $Y$ given by Eq. \eqref{eq:Y_od} if on top of the off-diagonal contribution to volatility one would add a fundamental contribution. The following discussion can thus be transposed to that case as well.}
\begin{equation} 
    \sigma^2 =  \frac{\sigma_F^2}{1 - Y^2 z_\infty^2}, 
\end{equation}
i.e. excess volatility induced by trading, independently of its information content. This is in line with the empirical results of \cite{wyart2008relation} (section 4), \cite{taranto2018linear} and \cite{bouchaud2018trades} (Figs. 13.2 and 14.5), and is of course related to the well-known excess volatility/excess trading puzzle, see e.g. \cite{shiller1981stock, odean1999investors, leroy2006excess} and \cite{bouchaud2018trades}, chapters 2 \& 20, see also \cite{gabaix2021search, bouchaud2010impact, van2024ponzi} for related discussions. 

Hence, even if the {\it deterministic}, decaying part of impact does not contribute to long-term volatility, its fluctuations might do the job. Of course, this somewhat contorted scenario relies on the assumption that the random component of impact has a permanent contribution to price changes, i.e. $z_\infty > 0$. Although this hypothesis is somewhat ad-hoc, the non-trivial result here is the relation between price impact and volatility given by 
\begin{equation}\label{eq:peak_F}
    \mathcal{I}(Q) \propto \sqrt{\frac{\sigma^2 - \sigma_F^2}{z_\infty^2}} \times \sqrt{\frac{Q}{\phi}}.
\end{equation}
If, on the other hand, the permanent contribution $z_\infty$ vanishes, then trivially $\sigma^2= \sigma_F^2$. This is the Efficient Market picture, where uninformed trades do not contribute to long-term volatility and our whole construction breaks down. However, this would leave $\theta_0$ undetermined and would not allow rationalizing why $\theta_0$ is proportional to $\sigma/\sqrt{\phi}$, as found empirically. Furthermore, the impact of uninformed metaorders (which is probably a large fraction of all metaorders) would generate large price reversion effects, which again are not observed.  

Quite interestingly, if the random component of impact (see Eq. \eqref{eq:price_dyn}) is assumed to be such that $\mathbb{E}[\varepsilon \eta]=0$, it will not contribute to the covariance between price changes and order imbalance considered in section \ref{sec:SQLvsAgg}. Indeed, by definition such a contribution to price changes does {\it not} contribute to the covariance between price changes and order imbalance. In such a scenario, only the fundamental component can contribute to this covariance, provided some metaorders are ``informed'', as we show in section \ref{subsec:pref_informed}.

\subsection*{Discussion}

We thus have three possible scenarios for generating long term volatility from metaorder impact alone. The idea that metaorders are correlated between one another, with roughly the same long memory as {\it within} each one of them, seems plausible, is compatible with available data and provides a natural extension of the propagator scenario: diffusive prices emerge from the subtle interplay between decaying impact and autocorrelated order flow. Such a scenario predicts a weak dependence of the prefactor $Y$ of the square-root law with the participation rate of the metaorder, as $\widetilde \varphi^\beta$ with $\beta \approx 0.2$.

The second scenario, which attributes long term volatility to the non-decaying impact of metaorders executed with large child orders does not seem very credible to us, because it would lead to substantial mean-reversion effects due to the impact decay of small child orders, which is at odds with empirical decay.

Finally, long term volatility may result from the random component of impact, assumed to be permanent. This bypasses the paradox that {\it average} impact decays and, in the absence of correlations between metaorders, should lead to subdiffusion. In such a scenario, volatility is induced by trading activity alone, even when average impact is zero. Still, the fact that impact fluctuations are proportional to average impact, as postulated in Eq. \eqref{eq:price_dyn}, is important to recover the correct relation between peak impact $\mathcal{I}(Q)$ and $\sigma \sqrt{Q/\phi}$, Eq. \eqref{eq:peak_F}, which is now independent of $\widetilde \varphi$.

In fact, the data shown in section \ref{sec:SQLvsAgg} strongly favours an interpretation based on the {\it average} impact of long-range correlated metaorders. The last scenario, where volatility arises from the {\it fluctuating} part of impact, does not pass the test, as it is unable to reproduce the detailed structure of the correlations between price changes and generalized volume imbalances.   

\end{document}